\documentclass{elsart}
\usepackage[dvips]{epsfig}
\usepackage{amssymb}
\usepackage{nicefrac}
\usepackage{natbib}
\usepackage{bm}
\newcommand{\lsim}{\raisebox{-0.6ex}{$\stackrel{{\displaystyle<}}{\sim}$}}
\newcommand{\gsim}{\raisebox{-0.6ex}{$\stackrel{{\displaystyle>}}{\sim}$}}

\renewcommand{\vec}[1]{\mbox{\boldmath$#1$}}
\newcommand{\Om}{\it{\Omega}}
\begin{document}
\begin{frontmatter}
\title{Geodynamo $\alpha$-effect derived from box simulations of rotating magnetoconvection}
\author{A. Giesecke},
\author{U. Ziegler},
\and
\author{G. R\"udiger}
\address{Astrophysikalisches Institut Potsdam, An der Sternwarte 16, D-14482 Potsdam, Germany}
\begin{abstract}
The equations for fully compressible rotating magnetoconvection are 
numerically solved in a Cartesian box assuming conditions roughly suitable for the
geodynamo. 
The mean electromotive force describing the generation of mean magnetic
flux by convective turbulence in the rotating fluid  
 is directly calculated from the simulations, and the corresponding
 $\alpha$-coefficients are derived. 
Due to the very weak density
 stratification the $\alpha$-effect changes its sign in the middle of the
 box. It is positive at the top and negative at the bottom of the convection
 zone. For strong magnetic fields we also find a clear downward advection of
 the mean magnetic field.  Both of the simulated effects have been predicted by
 quasi-linear computations \citep{1979PEPI...20..134S, 1992A&A...260..494K}. 
Finally, the  possible connection of the obtained profiles of the EMF with
mean-field models of oscillating $\alpha^2$-dynamos is discussed. 
\end{abstract}
\end{frontmatter}
%
%
%
%
%
%
\section{Introduction}
Convective motions in the fluid outer core influenced by rotation and
magnetic fields are able to maintain the Earth's magnetic field for long
times interrupted by occasionally occurring reversals of the dominating dipole component.  
The equations describing the physical processes in the fluid outer part of
the Earth's core are rather stiff
\citep{1995GAFD...79....1K,2000RvMP...72.1081R} showing a broad range of timescales on
which the characteristic behavior is observable.
The resulting and dominating advective timescale $\tau_{\mathrm{adv}}=d/u'$
leads to a very short coherence length of the physical quantities like velocity
or magnetic field which, as a consequence, requires a very high spatial
resolution in numerical simulations to include the effects of small-scale turbulence
\citep{2003GDS...31..181}.
In case of a
conducting and rotating fluid with large-scale density stratification a
convectively driven turbulence can generate a mean electromotive force (EMF)
parallel to the mean magnetic field --  a process known as $\alpha$-effect
\citep{1980mfmd.book.....K}.
But the Earth's fluid outer core, though a rapid rotator, is rather weakly
stratified so that it remains
unclear whether the amplitude of the $\alpha$-effect is sufficient to 
 maintain dynamo action.

Present-date global numerical simulations are unable to include small-scale
motions because of computational restrictions.
Nevertheless, they show quite satisfying results in the sense that they are
able to reproduce several features of the observed geomagnetic  field
\citep{1996PEPI...98..207G, 1999GeoJI.138..393C, 1999JCoPh.153...51K}. 
In virtue of these results it seems to be justified to ignore small-scale
fluctuations but, on the other hand, there still remain many open questions
when comparing such simulations with their restricted parameter range with
observational data (see e.g. \citeauthor*{2000GGG...1..62}, \citeyear{2000GGG...1..62}).
Background of our calculations is the idea to describe
a dynamo by an induction equation for the mean magnetic field
$\left<\vec{B}\right>$ with a prescribed productive term due to the
small-scale turbulence that ensures the
existence of dynamo action.

In mean-field dynamo theory the magnetic field and the velocity are split up in a mean part,
 $\left<\vec{B}\right>$ and $\left<\vec{u}\right>$, and a
fluctuating component, $\vec{B}'$ and $\vec{u}'$.
The time behavior
of the mean magnetic field $\left<\vec{B}\right>=\vec{B}-\vec{B'}$ is described by 
\begin{equation}
\partial_t\left<\vec{B}\right>=\nabla\!\times\!\left(\left<\vec{u}\right>\!\times\!\left<\vec{B}\right>\!+\!\vec{\mathcal{E}}\!-\!{\eta}\ \!\nabla\!\times\!\left<\vec{B}\right>\right)\label{mean_field_ind}
\end{equation}
with the mean electromotive force $\vec{\mathcal{E}}=\left<\vec{u}'\times\vec{B}'\right>$
which is usually expressed by  
\begin{equation}
{\mathcal{E}}_i = \left<\vec{u}'\times\vec{B}'\right>_i=
\alpha_{ij}\left<{B}_j\right>+\beta_{ijk}\partial_k\left<B_j\right>.
\label{eq_alpha}
\end{equation}
The tensor  $\alpha_{ij}$  correlates the
turbulent EMF due to small-scale motions with the large-scale magnetic field including the effects of
anisotropy.
The tensor $\beta_{ijk}$
is related to the turbulent diffusivity $\eta_T$ by $\beta_{ijk}=\eta_{\mathrm{T}}  \epsilon_{ijk}$.
The easiest way to look for dynamo action is to solve equation
(\ref{mean_field_ind}) with zero mean flow $\left<\vec{u}\right>$ and with an EMF taken
from equation (\ref{eq_alpha}) without turbulent
diffusivity ($\beta_{ijk}=0$).

Oscillating solutions for so-called $\alpha^2$-dynamos have recently been presented by
\citet{2003PhRvE..67b7302S}. 
They determined a  radial profile of a spherically symmetric and isotropic
$\alpha$ under the conditions that the 
dynamo mode with the lowest eigenvalue is an oscillating solution and all stationary modes (that usually
dominate $\alpha^2$-models) are damped. 
This constraint leads to an $\alpha$-effect with characteristic zeros in the  radial profile.
Calculations with a uniform radial  $\alpha$-coefficient have been
performed by \citet{2003A&A...406...15R} who, instead, included a latitudinal
variation of $\alpha$ and effects of anisotropy. 
They found oscillating $\alpha^2$-dynamos only for exotic exceptions.  

\citet{2001GApFD..94..263H} analyzed the behavior of a geodynamo-like
$\alpha\Om$-dynamo model which produces an axial dipole with reversals that
are induced by a fluctuating $\alpha$. The amplitude of the fundamental dipole
mode behaves as a damped particle under the influence of a random force in a bistable potential.
Here, in contrast to that, a configuration without any differential
rotation and, hence, without $\Om$-effect is considered.
The purpose of the present  paper is to calculate the turbulent EMF directly from
numerical solutions of the full set of nonlinear MHD-equations for a
convectively driven turbulent flow under the influence of rotation and subject
to an imposed magnetic field.
Our aim is to get a representation for the EMF based on simulations under conditions
characteristic for the geodynamo.
The results principally include effects of  
anisotropy of the $\alpha$-effect, its radial dependence and its
quenching properties. 
The derived $\alpha$-coefficients will serve as some input data for future mean-field
$\alpha^2$-dynamo
calculations.
A simplified geometry of a Cartesian box representing a small part of a
rotating spherical shell provides the ability to
examine the small-scale behavior of the fluid motions and magnetic field and
allows to consider effects of the turbulence that are neglected in
global simulations.

%
%
%
%
%
%
\section{The model}
\subsection{General properties}
\begin{figure}[htb]
\includegraphics[width=6cm]{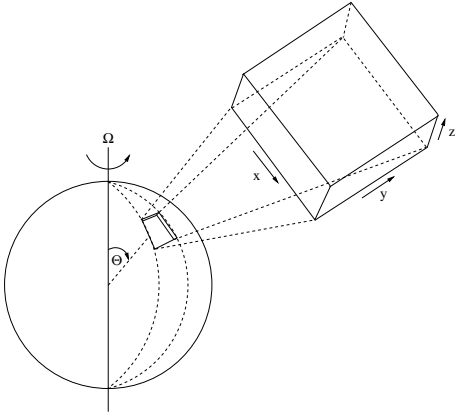}
\caption{ Model box being part of a rotating spherical shell at latitude
  $\theta$.} \label{modell_skizze}
\end{figure}

Our model is an adaption of configurations used, for example, by \citet{2001A&A...376..713O} and 
\citet{2002A&A...386..331Z}  who examined
rotating magnetoconvection in a box suitable for the solar convection zone.
Figure~\ref{modell_skizze} shows a sketch of the  box placed somewhere on a
spherical shell at some latitude $\theta$.
The coordinate system is chosen such that the unit vectors ${\vec{
\hat{x},\hat{y},\hat{z}}}$ form a
right-handed corotating system with $\vec{\hat{x}}$ pointing towards the
equator, $\vec{\hat{y}}$ pointing in the toroidal direction (from west to east) and $\vec{\hat{z}}$ pointing from the
bottom to the top of the box.
Translating this Cartesian system into global spherical coordinates, $\vec{\hat{z}}$
represents the radial direction $\vec{\hat{r}}$ directed from inside to
outside, $\vec{\hat{y}}$ the azimuthal direction $\vec{\hat{\phi}}$ and $\vec{\hat{x}}$ the
meridional direction $\vec{\hat{\theta}}$, respectively.
The angular velocity  $\vec{\Om}$ in the local box coordinate system 
is then given by
$\vec{\Om}=-\Om_0\sin\theta\hat{\vec{x}}+\Om_0\cos\theta\hat{\vec{z}}$
where $\Om_0$ is the angular velocity of the rotating spherical shell.

We try to construct a simple model that, 
at least, roughly represents the conditions  in the Earth's fluid 
interior. 
Our Cartesian box model consists of one convectively instable layer with a
weak density stratification.
The parameters have been chosen in a way that, ultimately,  a 
low Mach number flow with ${\rm Ma} \sim O(10^{-2})$ results and
compressibility effects, though existing, are rather small. 
We further restrict our computations to a rapidly  rotating box expressed
by a Rossby number ${\rm Ro}={u'/2\Om d} \ll 1$.
Our principal interest focuses on the presence of strong magnetic fields with significant
dynamical influence on the flow.
To investigate the transition from the
weak field case to the strong field case the strength of the imposed magnetic
field is successively increased covering a wide range of magnitudes.
Here, we present results obtained for a magnetic field applied in
$y$-direction corresponding to a toroidal field in spherical coordinates
i.e. the production of a poloidal field from a toroidal field via the
$\alpha$-coefficient $\alpha_{yy}$ is examined. 
Imposing other field components will be
the subject of subsequent studies.  
In the strong field case this configuration ensures that the inertial and viscous forces are
negligible and the main balance between the forces governing the
magnetoconvection state is given by the Coriolis force and the Lorentz
force as it is supposed to be the case in the
fluid core \citep{1996PEPI...98..163H,2000AnRFM..32..409Z}.
%
%
%
%
\subsection{Equations}
The MHD-equations for a rotating fluid including
the effects of thermal conduction, compressibility, viscous friction and
losses due to magnetic diffusivity are solved numerically using the code
NIRVANA \citep{1998CPC...109..111Z, 1999CPC...116..65Z}.
The equations in the local corotating system are 
\\
\begin{eqnarray}
\partial_t\rho&=&-\nabla\cdot(\rho\vec u) \label{conteq}\\
\partial_t(\rho\vec{u})&=&-\nabla\cdot(\rho\vec{uu})\!-\!\nabla\!
P\!+\!\nabla\!\cdot\!\sigma\!+\!\rho\vec{g}\!
-\!2\rho\vec{\Om}\!\times\!\vec{u}+\!\frac{1}{\mu_0}\!(\nabla\!\times\!\vec{B}\mathrm)\!\times\!\vec{B}\label{nseq}\\
\partial_t e&=&-\nabla\cdot(e\vec{u})-P\nabla\!\cdot\!\vec{u}+\sigma
\!\circ\!\nabla\vec{u}
+\frac{\eta}{\mu_0}|\nabla\!\times\!\vec{B}|^2+\nabla\!\cdot\!(\chi\nabla
T)\label{eneq}\\
\partial_t\vec{B}&=&\nabla\times(\vec{u} \times
\vec{B}-{\eta}\nabla\!\times\!\vec{B})\label{indeq}
\end{eqnarray}
with the density $\rho$, velocity $\vec{u}$, pressure $P$, magnetic flux
 density $\vec{B}$, temperature $T$ and the thermal energy density $e$. 
We assume a constant gravitational field
$\vec{g}=-g\hat{\vec{z}}$ within  the domain.
The viscous stress tensor $\sigma$ is given by
$\sigma_{ij}=\nu\rho\left(\partial{u}_{i}/\partial
  x_j\!+\!\partial{u}_{j}/\partial x_i\!-\!\nicefrac{2}{3}\nabla\!\cdot\!\vec{u}\
\delta_{ij}\right)$.
$\nu$ denotes the kinematic viscosity and $\chi$ the thermal
conductivity coefficient.
The values of $\chi$, the dynamic viscosity 
$\nu_{\mathrm{dyn}}=\nu\rho$ and the magnetic diffusivity $\eta$ are constant over the box volume.
An ideal gas equation of state is assumed with
$P=(\gamma-1)e={k}{(m\bar \mu)^{-1}}\rho T$ where $k$ is the Boltzmann constant, $m$ the atomic mass unit, 
$\bar\mu$ the
mean molecular weight ($\bar{\mu}=1$ for all runs) and
$\gamma=C_P/C_V=5/3$ is the ratio
of the specific heats.
The permeability $\mu_0$ is given by the vacuum value $\mu_0=4\pi\times 10^{-7}\mathrm{VsA^{-1}m^{-1}}$.
%
%
%
%
%
\subsection{The initial state}
From the equation of state
 and the condition for hydrostatic
equilibrium, $\partial_z P = -\rho g$, together with the assumption of a polytropic temperature distribution,
$T=T_0\left({\rho}/{\rho_0}\right)^{\Gamma}$, the initial density distribution can be calculated as
\begin{equation}
\rho(z) = \rho_0\left(1+\frac{\partial_zT}{T_0}(d-z)\right)^{1/\Gamma}
\label{eq_rhodistribution}
\end{equation}
where $d$ stands for the vertical box extension and the polytropic index $\Gamma$
is given by
$\Gamma=\ln \left(1+d\partial_zT/T_0\right)/\ln \xi$.
The stratification index $\xi =\rho_{\mathrm{bot}}/\rho_{\mathrm{top}}$,
the temperature $T_{\mathrm{0}}$ and the global temperature gradient $\partial_{{z}}T$ are prescribed
input parameters whose values are given below. The subscript 0 refers to values taken at the top boundary of the box.

The gravitational acceleration can be calculated from the hydrostatic
equilibrium condition and the initial density distribution (\ref{eq_rhodistribution})
and is given by
\begin{equation}
g =\frac{\Gamma+1}{\Gamma}\frac{k}{m\bar \mu}\partial_z T.
\label{g}
\end{equation}
To obtain a convectively unstable state the condition $\Gamma > \gamma-1$
must be fulfilled.
In fact, $\Gamma=\gamma-1$ leads to a Rayleigh number
\begin{equation}
{\rm Ra}=\frac{\rho{g C_P}d^4}{\chi\nu}T\left({\partial_{z}T}-\frac{g}{C_{P}}\right)=0
\label{gg}
\end{equation}
 with $C_P={k}({m\bar{\mu}})^{-1}\gamma(\gamma-1)^{-1}$ the specific heat at
constant pressure.
Here, parameters are chosen such that $\Gamma>\gamma-1$.
%
%
%
%
\subsection{Boundary Conditions}
All quantities are subject to periodic boundary conditions in the horizontal directions.
At the top and at the bottom of the computational
domain constant values for density and temperature are imposed.
The vertical boundary condition for the magnetic field is a perfect conductor condition, and a stress-free
boundary condition is adopted for the horizontal components of the velocity $u_x$ and $u_y$.
Impermeable box walls at the top and at the bottom lead to a vanishing
$u_z$ at the vertical boundaries. 
Table~\ref{bc_tab} summarizes these conditions and gives the
initial values for density and temperature which describe the overall stratification and the global temperature
gradient.
\begin{table}[!h]
\caption{Vertical boundary conditions}
\begin{tabular}{lllll}
\hline
 & $\rho$  & $T$ & $\vec{u}$ &
$\vec{B}$
\\
\hline
{ top} & $1$ & $1$ & $\partial_z u_x = 0$ & $\partial_z B_x = 0$
\\
 $(z=d)$ & & & $\partial_z u_y = 0$ &  $\partial_z B_y = 0$
\\ 
& &  & $u_{z} = 0$ & $B_{z} = 0$
\\
\hline
\hline
{bottom}  & $1.1$ & $2$ & $\partial_z u_x = 0$ & $\partial_z B_x = 0$
\\
$(z=0)$ & & & $\partial_z u_y = 0$ & $\partial_z B_y = 0$
\\ 
& & & $u_{z} = 0$ & $B_{z} = 0$
\\
\hline
\end{tabular}
\label{bc_tab}
\end{table}
%
%
%
%
\subsection{Input parameters}
All input quantities are measured at the top of
the box.
This makes sense since the density variation with depth is negligible and the temperature
varies only by a factor of 2.
The parameters $\nu, \chi,\eta$ and $\Om$ are calculated from the Rayleigh
number Ra as defined above, the Prandtl number ${\rm Pr}=\nu\rho
C_{{P}}/\chi$, the magnetic
Prandtl number ${\rm Pm}=\nu/\eta$
and the Taylor number ${\rm Ta}={4{\Om}^2d^4}/{\nu^2}$.  
The basic parameter set used for all
simulations that are presented in this paper is given by ${\rm Ra}=10^6, {\rm
Pr}=0.5, {\rm Pm}=0.5$ and
${\rm Ta}=10^7$.

The Els\"asser number 
\begin{equation}
\Lambda={\vec{B}^2\over{2\Om\mu_0\rho\eta}}
\label{els}
\end{equation}
serves as an input
parameter for the magnitude of the imposed magnetic field whose influence is
investigated by varying ${\Lambda}$ from $10^{-2}$ 
to $10^3$
covering the
full range from weak fields to very strong fields.
The box with an aspect ratio 8:8:1 is placed at a latitudinal angle of
$45^{\circ}$ on the northern hemisphere of the rotating spherical shell and  a
standard resolution of $100\times100\times80$ grid points is used in all calculations. 
For all simulations temperature and density at the top of the box are
scaled to unity, as it is the case for the
global temperature gradient $\partial_z T$ and the box height $d$. 
A stratification index of $\xi=1.1$ is used.
%
%
%
%
%
%
\section{Results}
\subsection{General properties and energetics}
At first, a non-rotating non-magnetic
convection model is computed and the resulting statistically steady state
is used as initial condition for the full
problem of rotating magnetoconvection. 
Typical values for the turbulent velocity $\vec{u}'$ and the
turbulent magnetic field $\vec{B}'$ are obtained by an averaging procedure that includes the whole box
volume.
In the following, volume averages are indicated by double brackets, $\left<\left<\cdot\right>\right>$, whereas
horizontal averages are denoted by single brackets, $\left<\cdot\right>$.
As root mean square value of fluctuations we define 
$\left<\left<{f}'^2\right>\right>=(N_x N_y N_z)^{-1}\sum\limits_{i,j,k}
\left({f_{ijk}}-\left<{f}\right>_k\right)^2$.
Here, $N_x (N_y, N_z)$ denotes the number of grid cells in $x, (y, z)$
direction and
${f_{ijk}}-\left<{f}\right>_k$ is the deviation of the
fluctuating quantity at a certain grid cell labeled $ijk$ from its horizontal average.
Note that due to the horizontal averaging procedure and the periodic horizontal boundary
conditions the mean
quantities have no dependence on $x$ or $y$.   

Time averages are labeled by $\overline{f}$ and are computed only over time intervals that show no significant
change in the average itself.
In the following, time is measured relative to the turnover time given by 
$\tau_{\mathrm{adv}} = d/u_{\mathrm{rms}}$, where $u_{\mathrm{rms}}=\overline{\sqrt{\left<\left<{\vec{u'}^2}\right>\right>}}$.
All time averages are calculated over a time range of at least $20\tau_{\mathrm{adv}}$ starting at a certain time after
the effects of magnetic field and rotation have been introduced and $\left<\left<\vec{u'}^2\right>\right>$ has reached a new statistically steady
state.
A comparison with long-term computations shows that this is a
sufficient timespan in order to obtain meaningful results.
The longest run has been performed for $\Lambda=4$ -- namely more
than 110 turnover times -- which corresponds to about two magnetic
diffusion times $\tau_{\eta}=d^2/\eta$.
Note that $\tau_{\mathrm{adv}}$ evolves as a part of the solution and depends on
the imposed magnetic field whereas $\tau_{\eta}$ can be obtained from the
input parameters. 

\begin{figure}[htb]
\includegraphics[width=6.5cm]{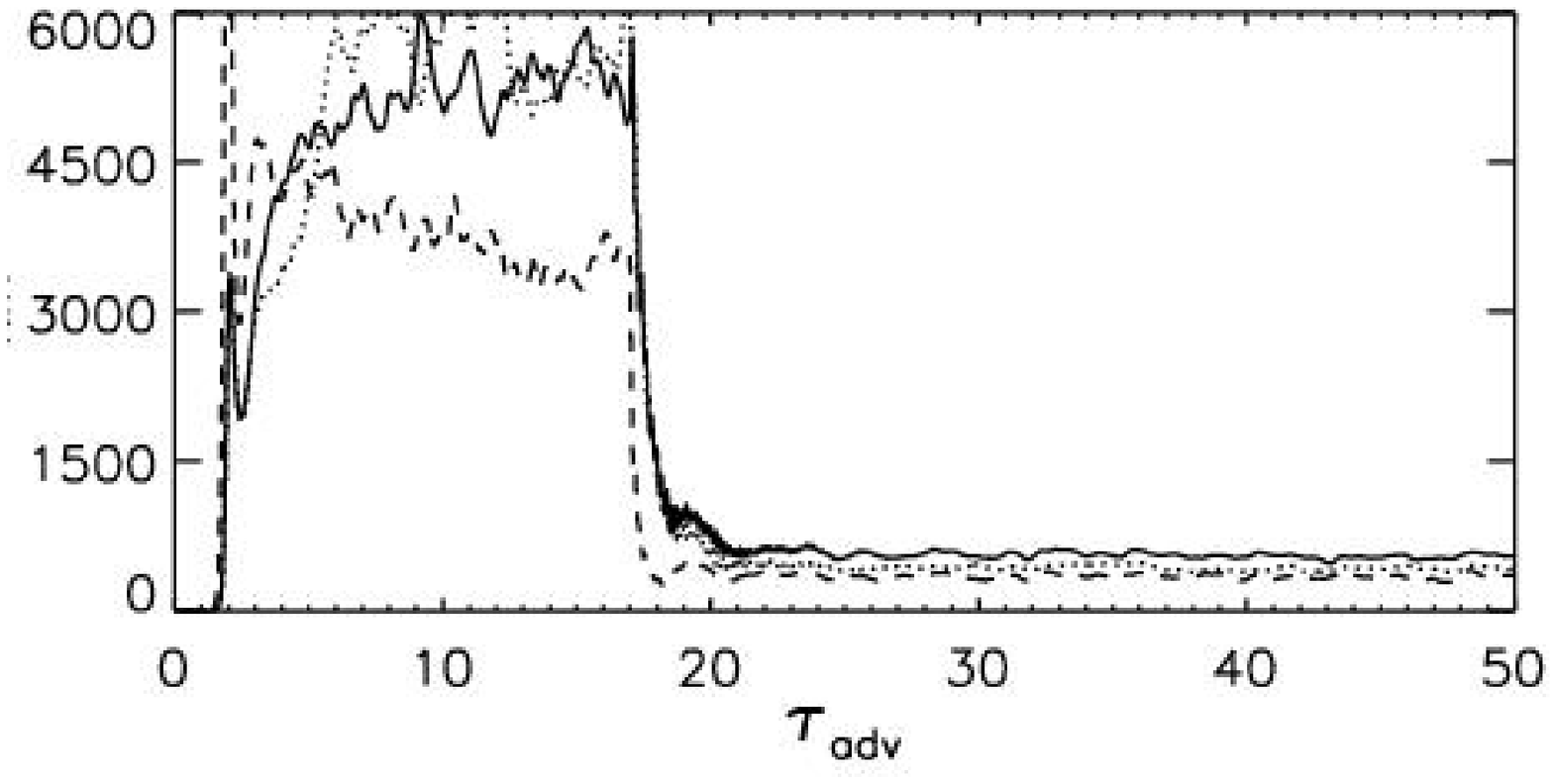}
\includegraphics[width=6.5cm]{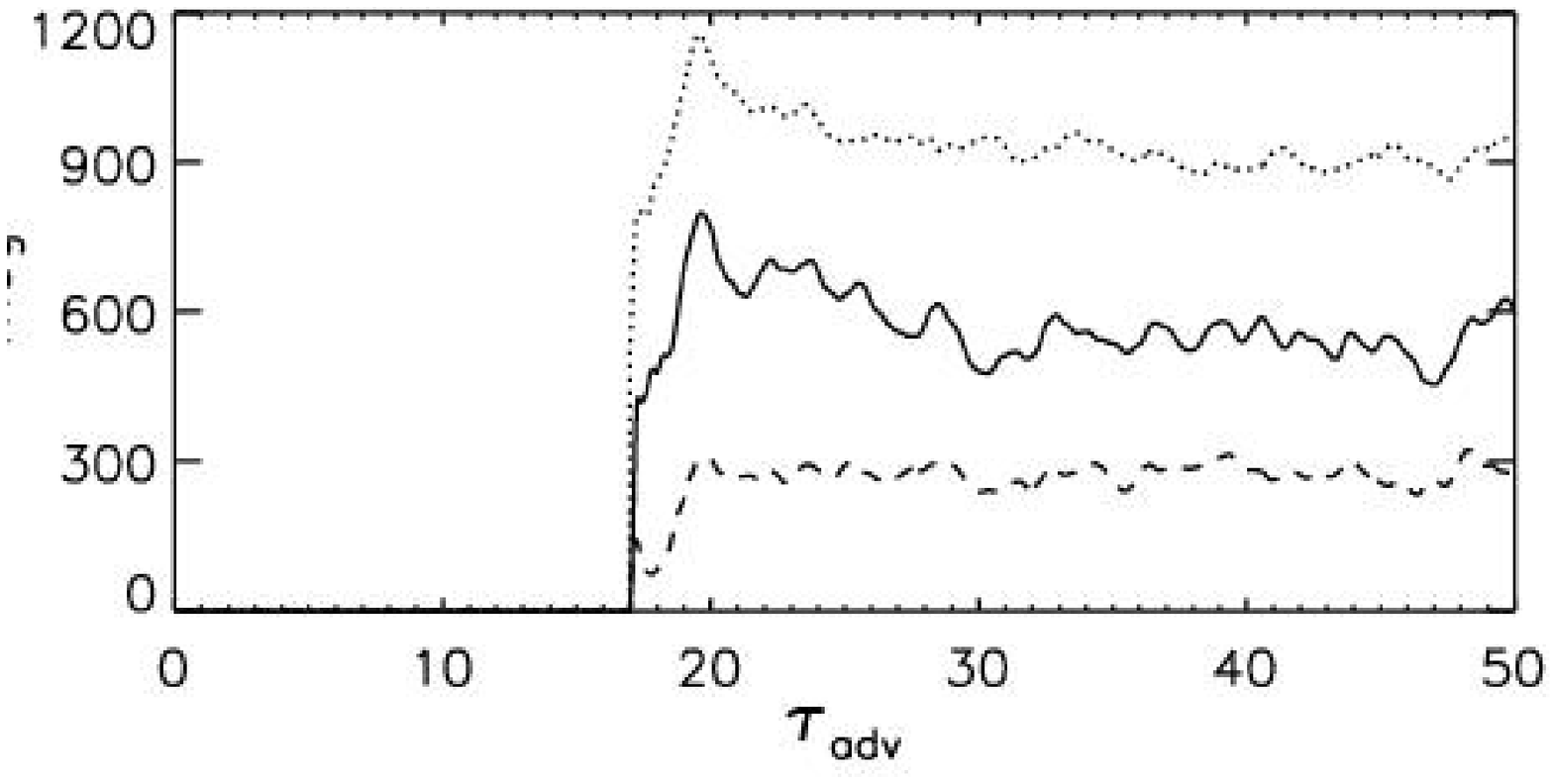}
\caption{\label{ekin/emag} Time dependence of the kinetic energy and the magnetic energy. $x$-component (solid), 
$y$-component (dotted), $z$-component (dashed). $\Lambda=1$.} 
\end{figure}

Figure~\ref{ekin/emag} shows the temporal behavior of the kinetic energies
$E_{\mathrm{kin}}^{x,y,z}=\int dV\rho
u^2_{x,y,z}/2$
(left) and the magnetic energies $E_{\mathrm{mag}}^{x,y,z}=\int dV(2\mu_0)^{-1}
B^2_{x,y,z}$ (right) for the simulation
 with $\Lambda=1$.
The first part of the simulation up to $t\approx 17\tau_{\mathrm{adv}}$
corresponds to 
thermal convection without rotation and without magnetic field.
At $t\approx 17\tau_{\mathrm{adv}}$  the effects of rotation and magnetic field are
added seen by a significant drop in the kinetic energies (left
panel of figure~\ref{ekin/emag}) and
a sharp increase of magnetic energies (right
panel of figure~\ref{ekin/emag}). 
After a short transition phase the energy of the different components remains approximately constant
for the remainder of time. 
A large-scale magnetic field in $x$-direction establishes during the timespan
$t=17\tau_{\mathrm{adv}}...20\tau_{\mathrm{adv}}$ (see also Figure~\ref{meanbfield} below) which
is associated with a remarkable amount of
magnetic energy stored in the $x$-component
(solid line in the right panel of figure~\ref{ekin/emag}).  

Since the induction equation (\ref{mean_field_ind}) for 
the mean magnetic field gives
$\partial_t\left<B_z\right>=0$ no $\left<B_z\right>$ can evolve during the simulations.  
Therefore, the 
magnetic energy in the vertical field component results from the  fluctuating
component $B_z'$.
The time dependence of the energies is qualitatively similar for all runs.
Major differences appear in the amount of quenching after 
the effects of rotation and magnetic field have abruptly been introduced.

\begin{figure}[htb]
\includegraphics[width=7cm]{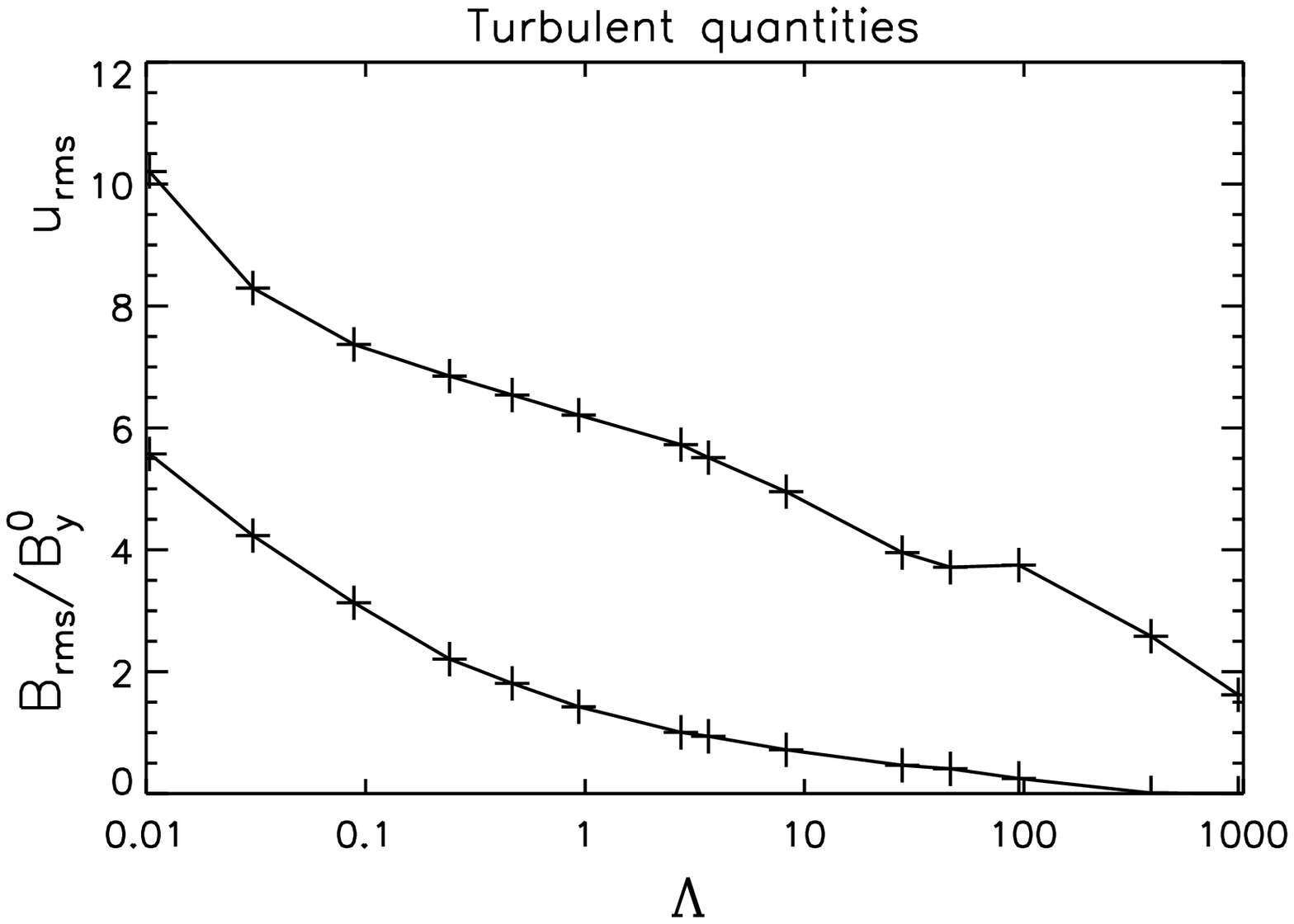}
\hspace*{0.5cm}
\includegraphics[width=7cm]{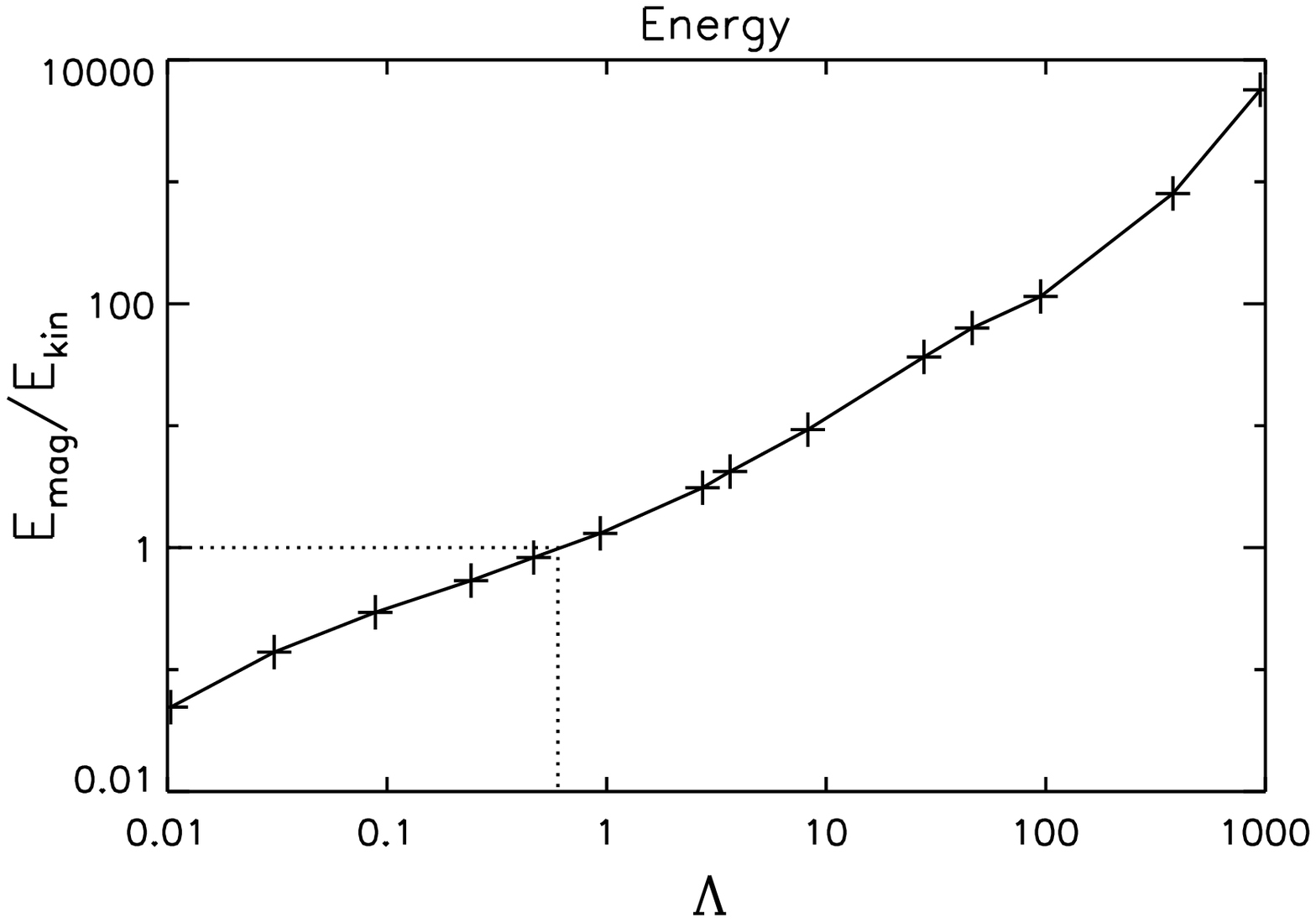}
\caption{\label{usbs}  Left: Quenching of turbulent velocity and (normalized) magnetic field fluctuations.
 Right: Ratio of the magnetic energy to the kinetic energy as a function of $\Lambda$.} 
\end{figure}

The behavior of the fluctuating quantities is shown in Figure~\ref{usbs} where the turbulent velocity
${u_{\mathrm{rms}}}$ and the normalized turbulent magnetic field
${B}_{\mathrm{rms}}/B_y^0$ (where $B_{\mathrm{rms}}$ is defined analog to $u_{\mathrm{rms}}$) in
dependence of the imposed magnetic field respective $\Lambda$ is presented.
Both quantities are significantly reduced compared to a non-magnetic but 
rotating convective state, indicating the trend to a more laminar flow.
%
%

The ratio between the total magnetic energy and the kinetic
energy is plotted in Fig.~{\ref{usbs}} (right). 
Equipartition is reached for $\Lambda \approx 0.6$ indicated by the dotted lines.
For $\Lambda>0.6$ the magnetic energy clearly exceeds the kinetic energy. 
For $\Lambda = 100$ the magnetic energy dominates the kinetic energy by a factor of 100.

When increasing $\Lambda$ from 0.01 to 1000 the combined effects of
rotation and magnetic field lead to Rossby numbers from
$7\cdot10^{-2}$ to $1\cdot10^{-2}$ and turbulent Mach numbers  $\mathrm{Ma}={u_{\mathrm{rms}}}/c_s$ from $7\cdot10^{-2}$ to $1\cdot 10^{-2}$.
This is still much larger than the real Mach number in the fluid outer
core which is assumed to be of the order of $10^{-7}$.
%
%
Reynolds numbers,  
$\mathrm{Re}={u_{\mathrm{rms}}}d/\nu$, are in the range from Re=210 for $\Lambda
=0.01$ and Re=32 for $\Lambda=1000$.
%
%
%
%
\subsection{Patterns of flow and magnetic field}
\begin{figure}[htb]
\includegraphics[width=6.5cm]{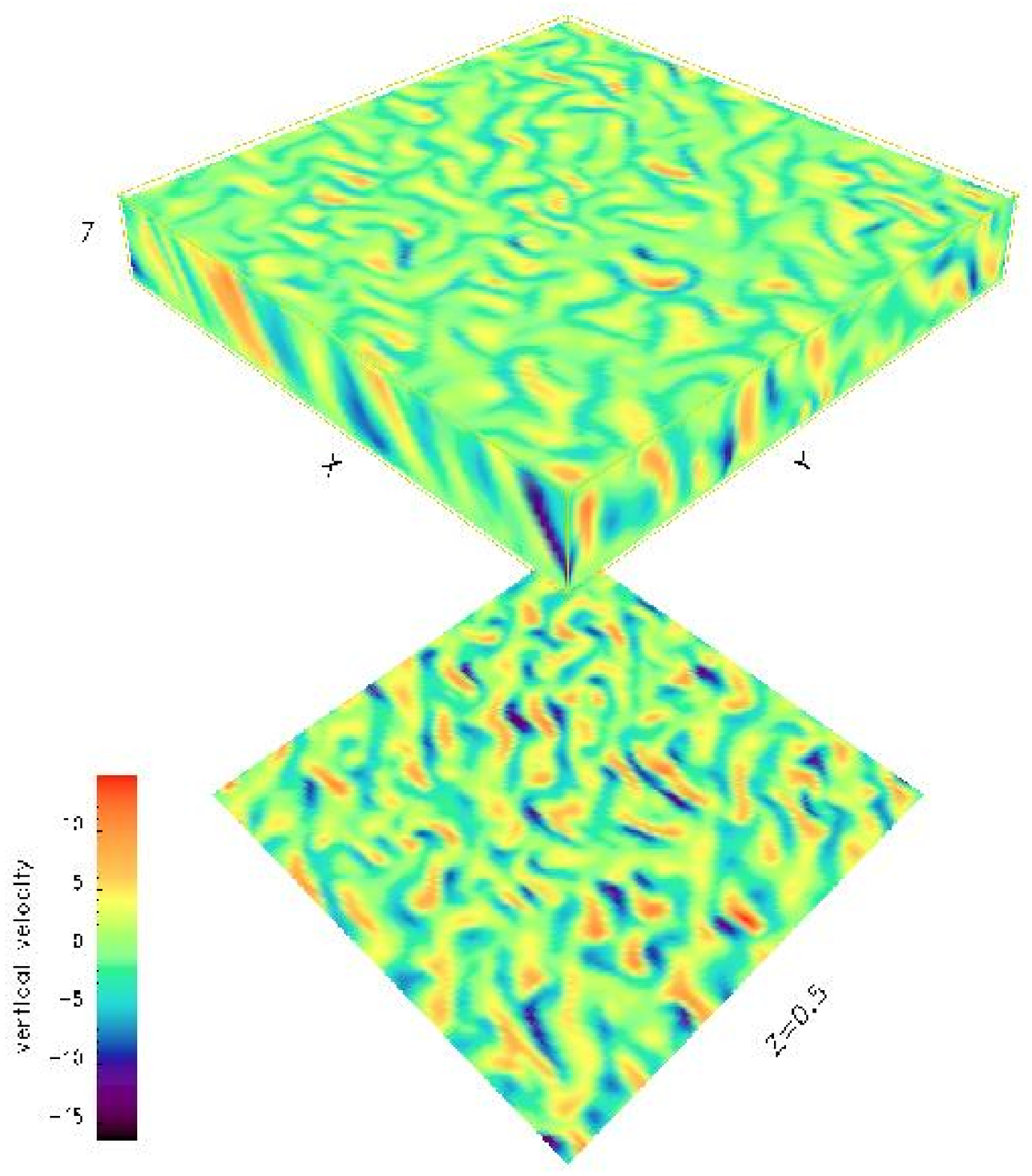}
\includegraphics[width=6.5cm]{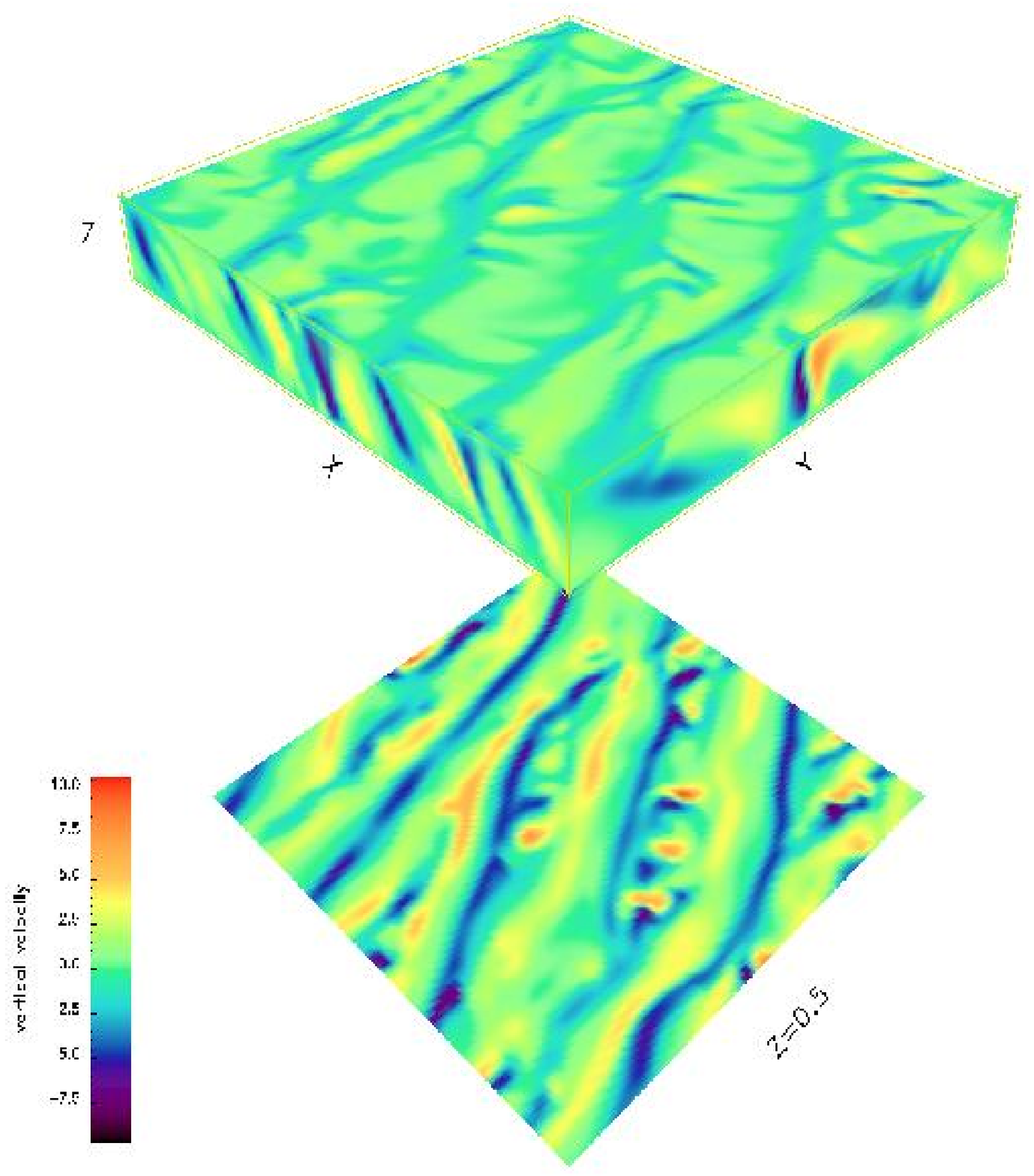}
\caption{\label{vz3Dpattern} Vertical velocity pattern of rotating
  magnetoconvection for $\Lambda=1$ (left) and   $\Lambda=10$ (right).}
\end{figure}

The dynamical influence of the imposed magnetic field can be seen in
figure~\ref{vz3Dpattern} where the $z$-component of the velocity near the domain faces and in a
horizontal plane at $z=0.5$ is shown.
The left (right) panel shows a snapshot of the developed rotating magnetoconvection
 for $\Lambda\approx 1$ (10).
Upflows are visualized in grey whereas downflows are represented in dark tones.
Compared to non-magnetic, rotating convection the magnetoconvection pattern remains nearly unchanged for $\Lambda
\lsim 1$.
Many topologically not connected columnar  convection cells  can be seen tilted by an
angle of $45^{\circ}$ with respect to the $z$-axis and become aligned with the
rotation axis (Taylor-Proudman theorem).

Stronger magnetic fields lead to remarkable changes
in the flow pattern.
Between $\Lambda=1$ and $\Lambda=4$ the quasi-regular pattern becomes more and more
disintegrated and evolves towards a nearly two-dimensional flow as illustrated in the right panel of Figure~\ref{vz3Dpattern}.
The convection cells are clearly elongated
along the imposed magnetic field direction ($y$-direction) and show little
variations of the convective
velocity along the field lines.
The sheetlike convection cells are again tilted inside the box and aligned
with the rotation axis $\vec{\Om}$ as it is the case for the weak-field calculations. 
Compared to the cases of rotating convection or weak-field rotating
magnetoconvection, the strong-field case is further characterized by a
significant reduction of the number of convective cells.
The nearly two-dimensionality of the flow can be explained from a condition similar
to the Taylor-Proudman theorem for rotating spheres:
For a stationary state with small
deviations from the basic unperturbed state and neglecting diffusive
terms it follows from the induction equation that $B_i\partial_i u_j=0$ i.e. 
motions cannot vary in the direction of the imposed magnetic field \citep{1961hhs..book.....C}.
%
%
%
%
\subsection{Mean fields and $\alpha$-coefficients}
\begin{figure}[htb]
\includegraphics[width=4.5cm]{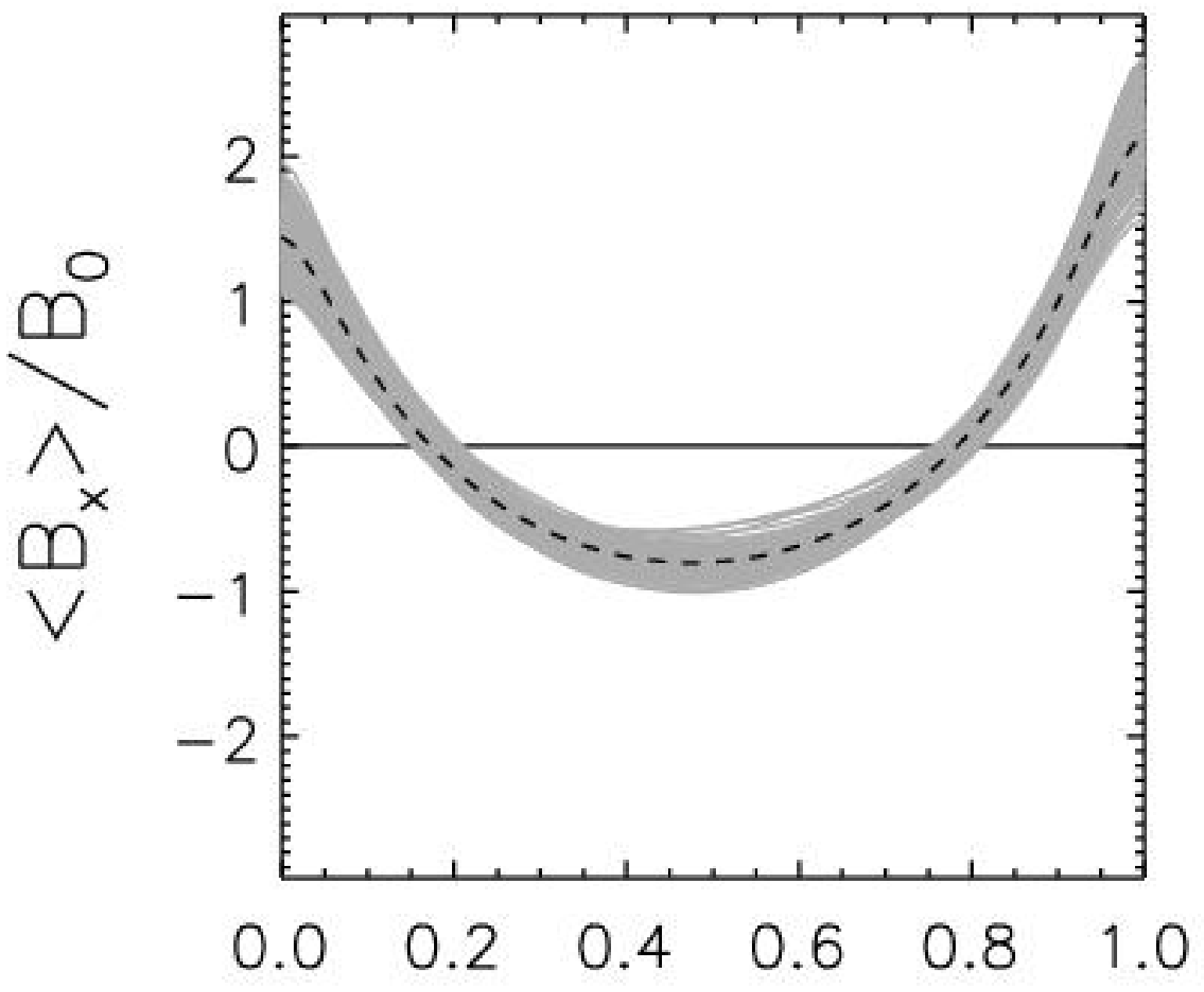}
\includegraphics[width=4.5cm]{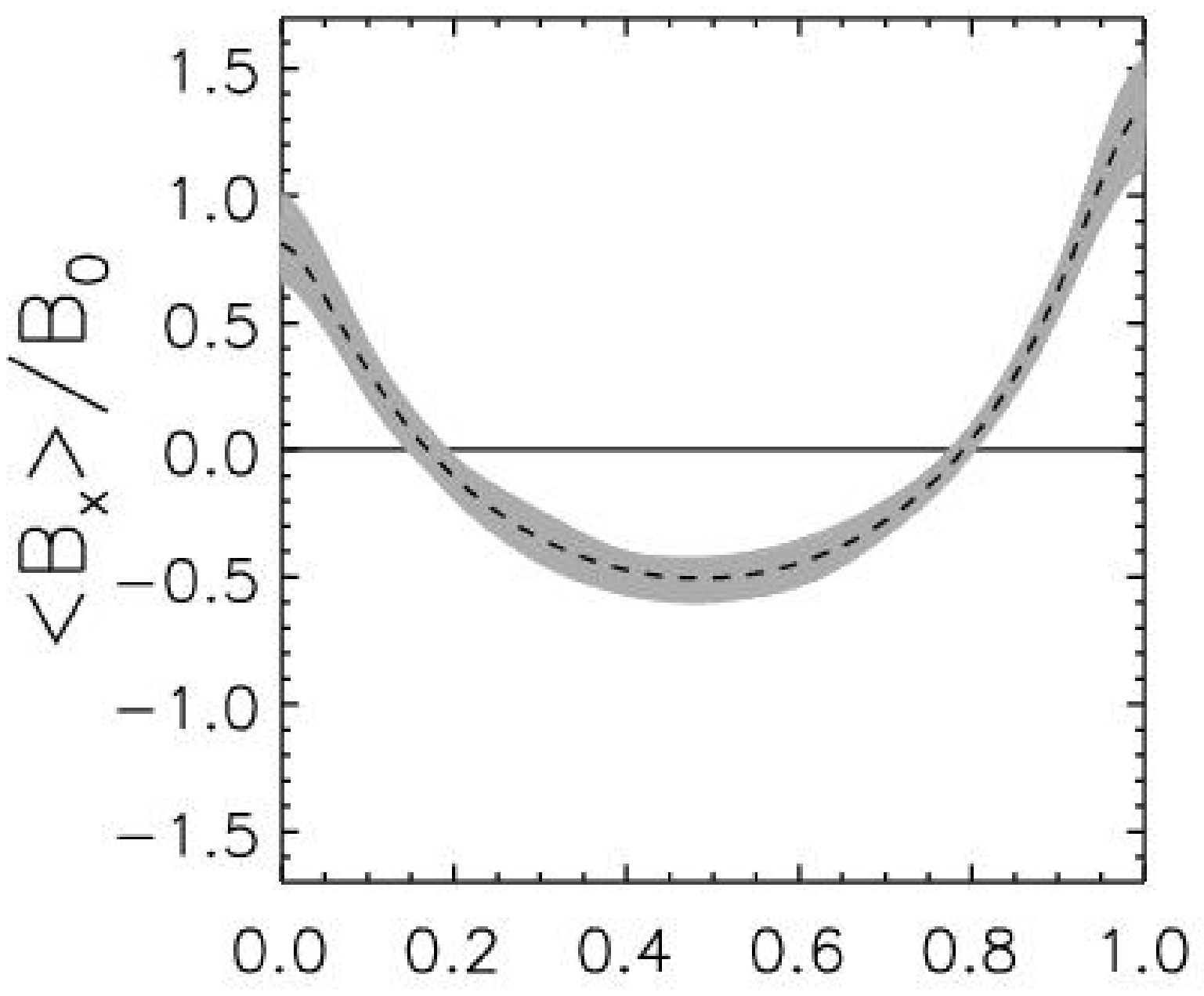}
\includegraphics[width=4.5cm]{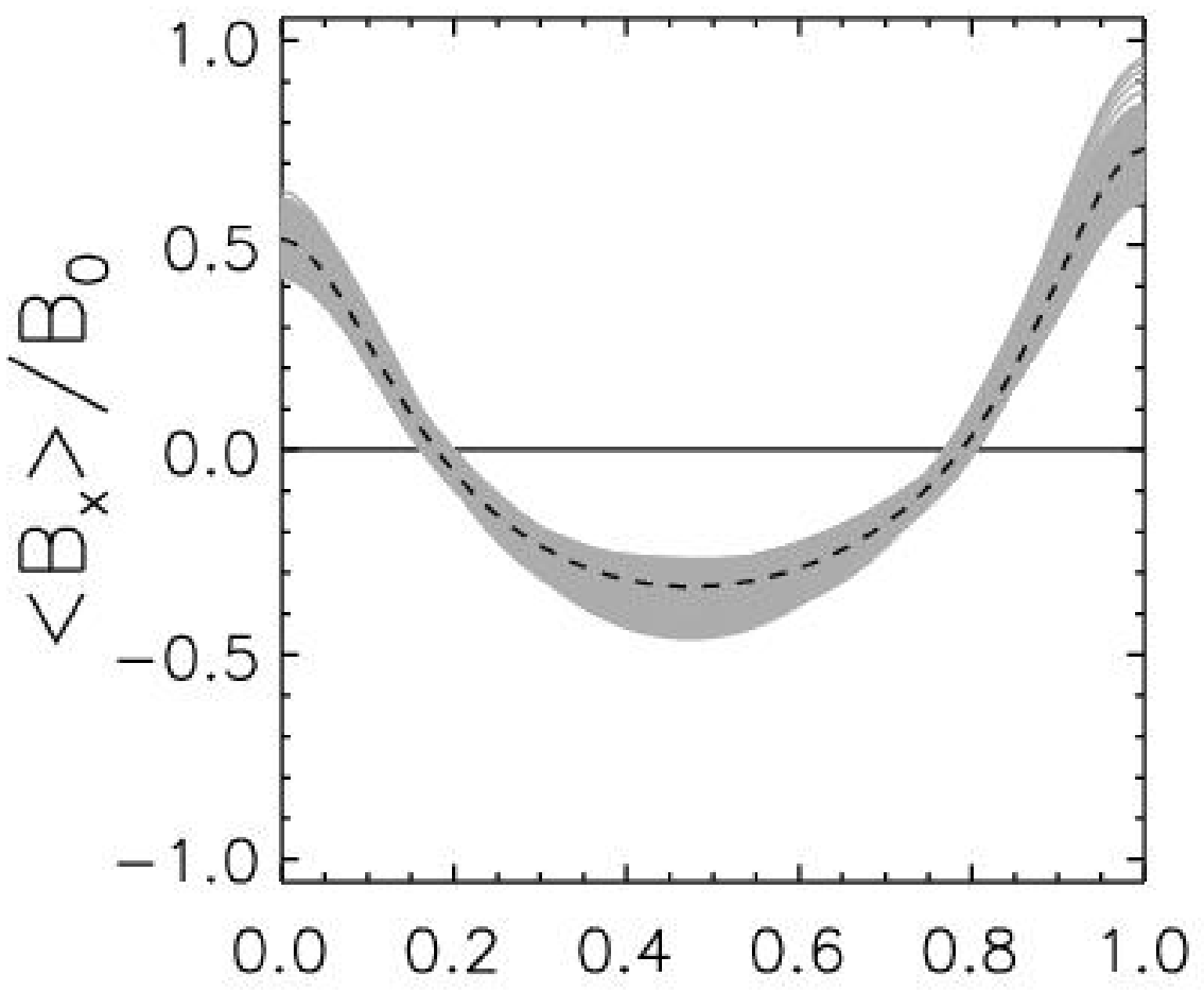}
\\
\includegraphics[width=4.5cm]{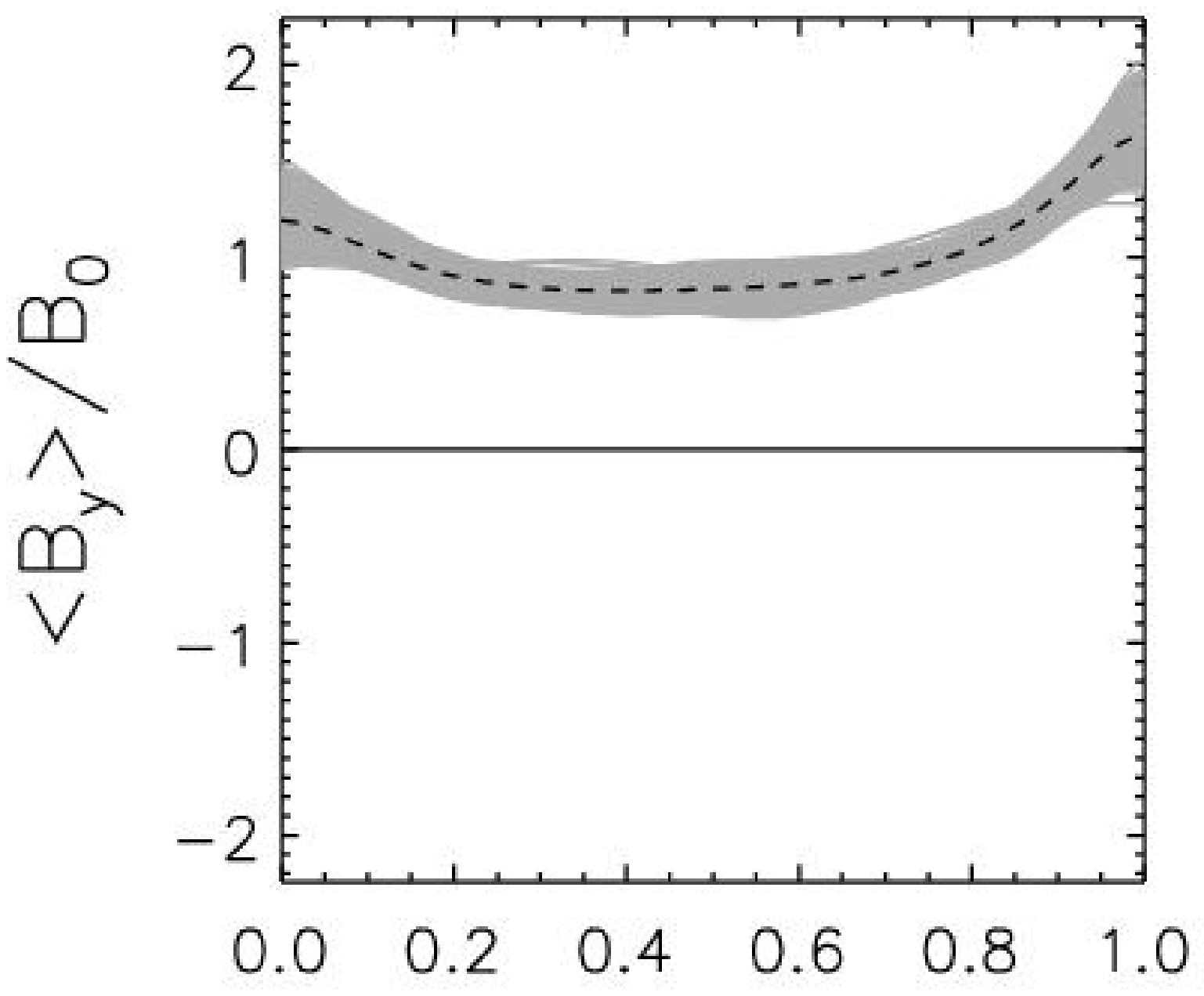}
\includegraphics[width=4.5cm]{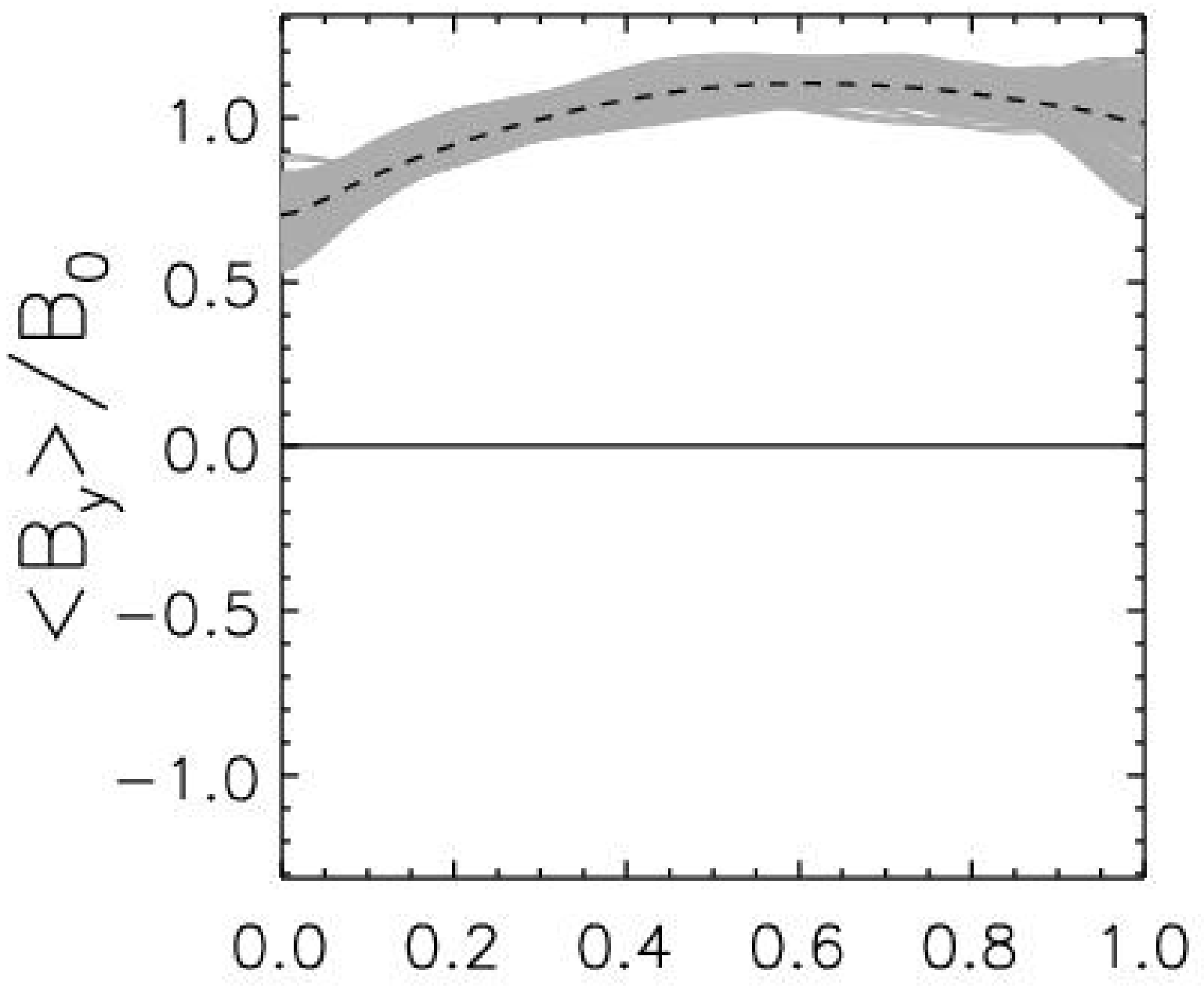}
\includegraphics[width=4.5cm]{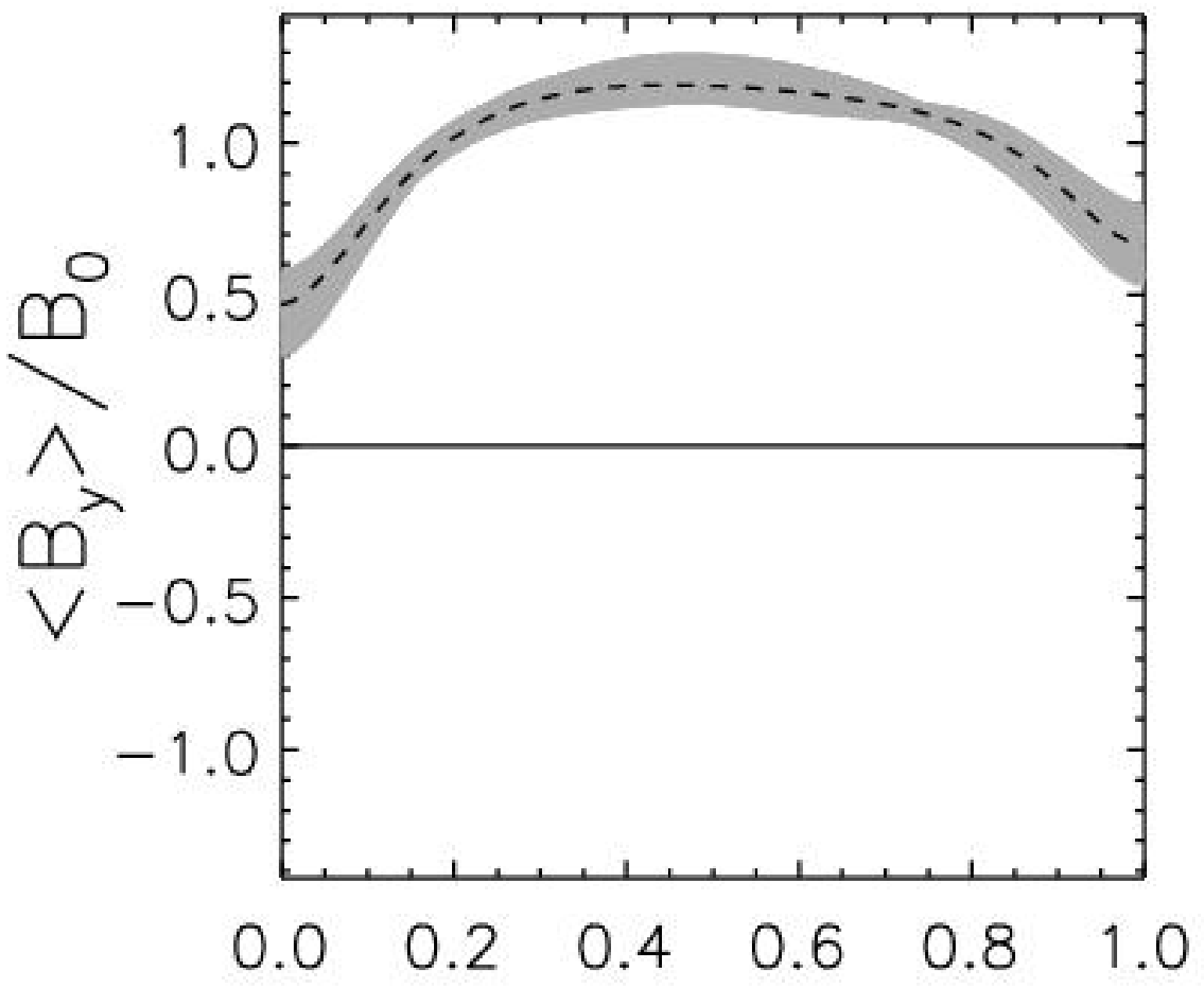}
\\
\includegraphics[width=4.5cm]{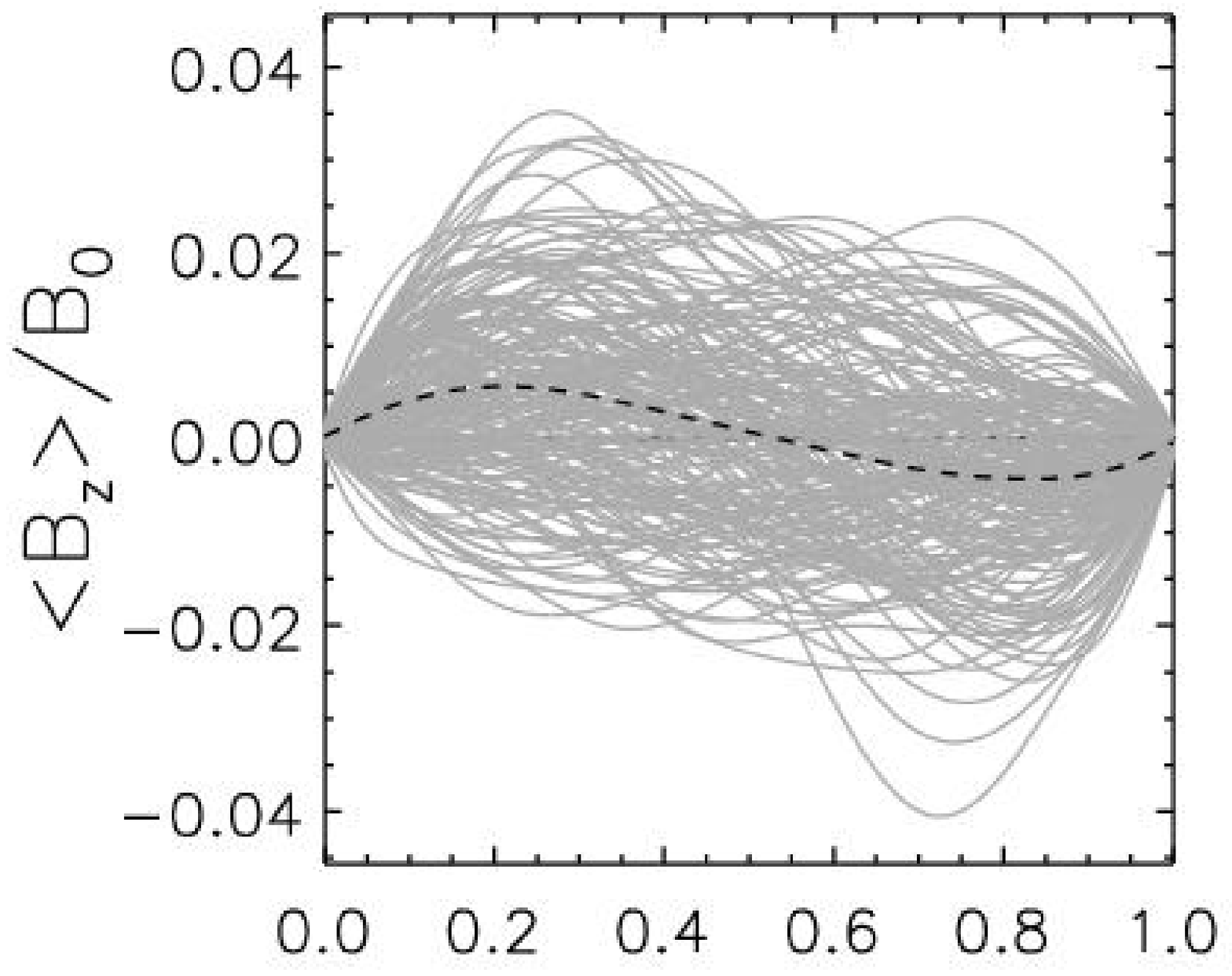}
\includegraphics[width=4.5cm]{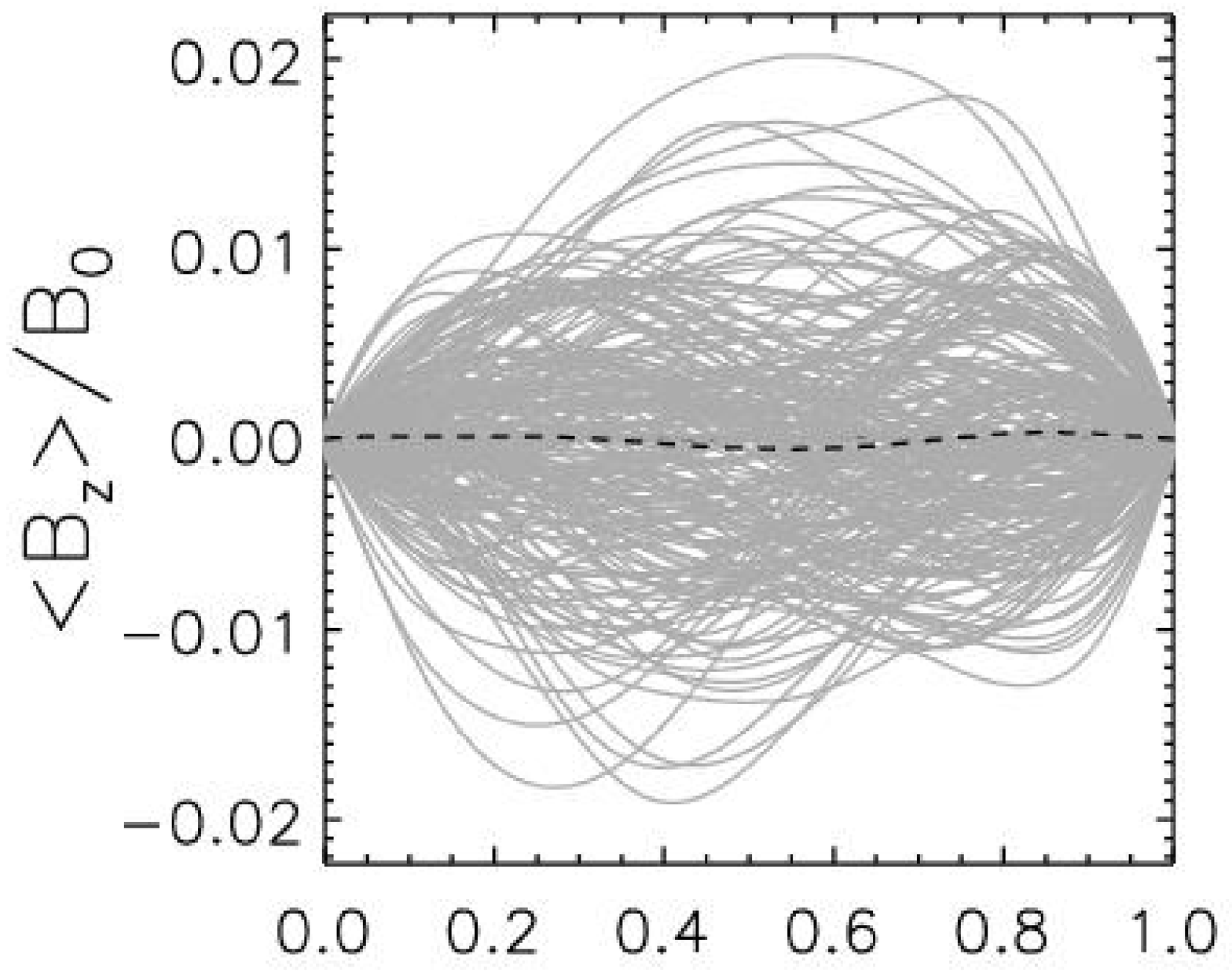}
\includegraphics[width=4.5cm]{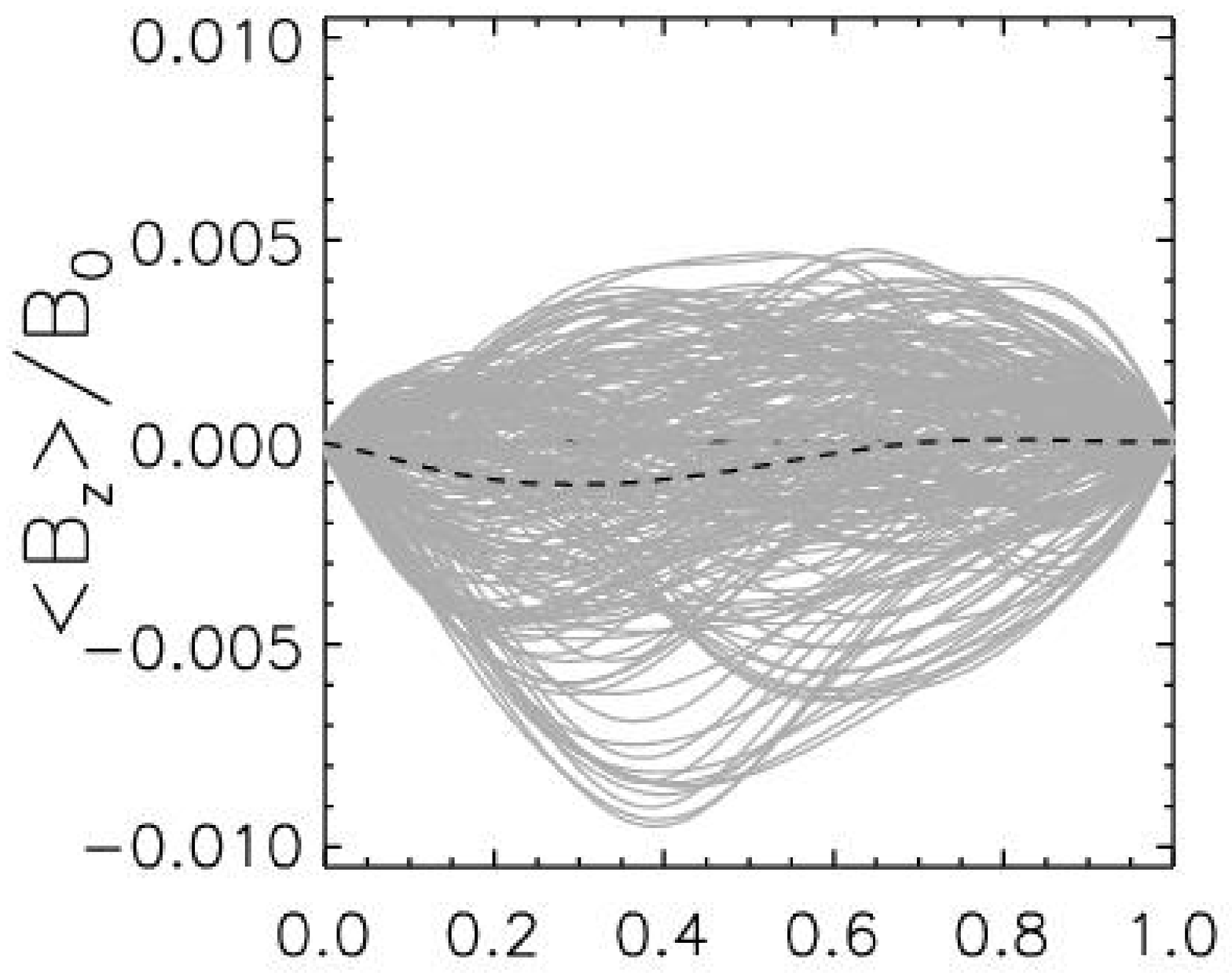}
\\
\caption{$z$-dependence of mean magnetic fields for $\Lambda = 0.1, 1, 10$ (from left to right). From top to bottom: $\left<B_x\right>,
\left<B_y\right>$ and $\left<B_z\right>$}
\label{meanbfield}
\end{figure}

Figure~\ref{meanbfield} shows the components of the horizontally averaged magnetic
field in
units of the initial field $B_0$ for the cases $\Lambda=0.1,1,10$. 
The light gray lines
represent different snapshots within the averaging period indicating
substantial fluctuations, and the thick dashed line gives the time average of these individual curves. 
There exists in all cases a significant magnetic field component
$\left<B_x\right>$ close to the boundaries and in the bulk having opposite
signs in that regions. 
This behavior strongly differs from the results reported by
\cite{2002A&A...386..331Z} for highly stratified convection where the $x$-component of the magnetic field
was found negligible ($\left<B_x\right> \ll
\left<B_y\right>$).
This is probably due to a much lower Taylor number employed in the simulations
of \cite{2002A&A...386..331Z} and due to the initial two-layer configuration consisting of a
convectively instable layer on top of a stable layer.
The generation of $\left<B_x\right>$ at the vertical boundaries is a result of the
perfect conductor boundary conditions for the magnetic field which means that no magnetic flux can cross the 
boundaries favoring concentration of magnetic flux at the top and at the bottom of
the domain.

As 
expected, no significant mean magnetic field in $z$-direction establishes.
For weak imposed fields local dynamo action leads to a
slight amplification of $\left<B_y\right>$
close to the boundaries.
This effect does not take place for stronger imposed fields where, in contrast,
a reduction of $\left<B_y\right>$ occurs at the vertical boundaries.

Usually, the dynamo $\alpha$-coefficients are computed in a simplified way by
\begin{equation}
\mathcal{E}_x=\alpha_{xy}\left<B_y\right>,\qquad\mathcal{E}_y=\alpha_{yy}\left<B_y\right>,\qquad\mathcal{E}_z=\alpha_{zy}\left<B_y\right>.
\label{alpha}
\end{equation}
This may not be justified with the mean field configuration obtained here, since the
presence of a $\left<B_x\right>$ and the presence of gradients of the mean
field components close
to the vertical boundaries lead to a contribution from non-diagonal
coefficients of the $\alpha$-tensor (describing the anisotropy) and from
turbulent diffusivity.
Their contribution cannot be easily separated without contradiction, so it seems more
reasonable to regard  the calculated EMF-components as the more
relevant quantities.
This is  confirmed  by  calculations of
\cite{1990A&A...232..277B} who showed that the expressions
(\ref{alpha}) together with the mean field induction equation
(\ref{mean_field_ind}) are too simple for reproducing the
mean-field components as computed numerically from solving the full set of MHD-equations. 
However, these workers have assumed that this deviation occurs for shorter
timescales whereas for longer timescales the correct time dependence of the mean-field is retained if
a suitable quenching function for $\alpha$ is used. 
Due to the well-known theoretical background of characteristic properties for
the $\alpha$-coefficients and to provide input data for a mean-field
$\alpha^2$-dynamo model we nevertheless use relationship (\ref{alpha}) keeping
in mind that it stands for a more or less rough approximation.

\begin{figure}[h]
\includegraphics[width=3.4cm]{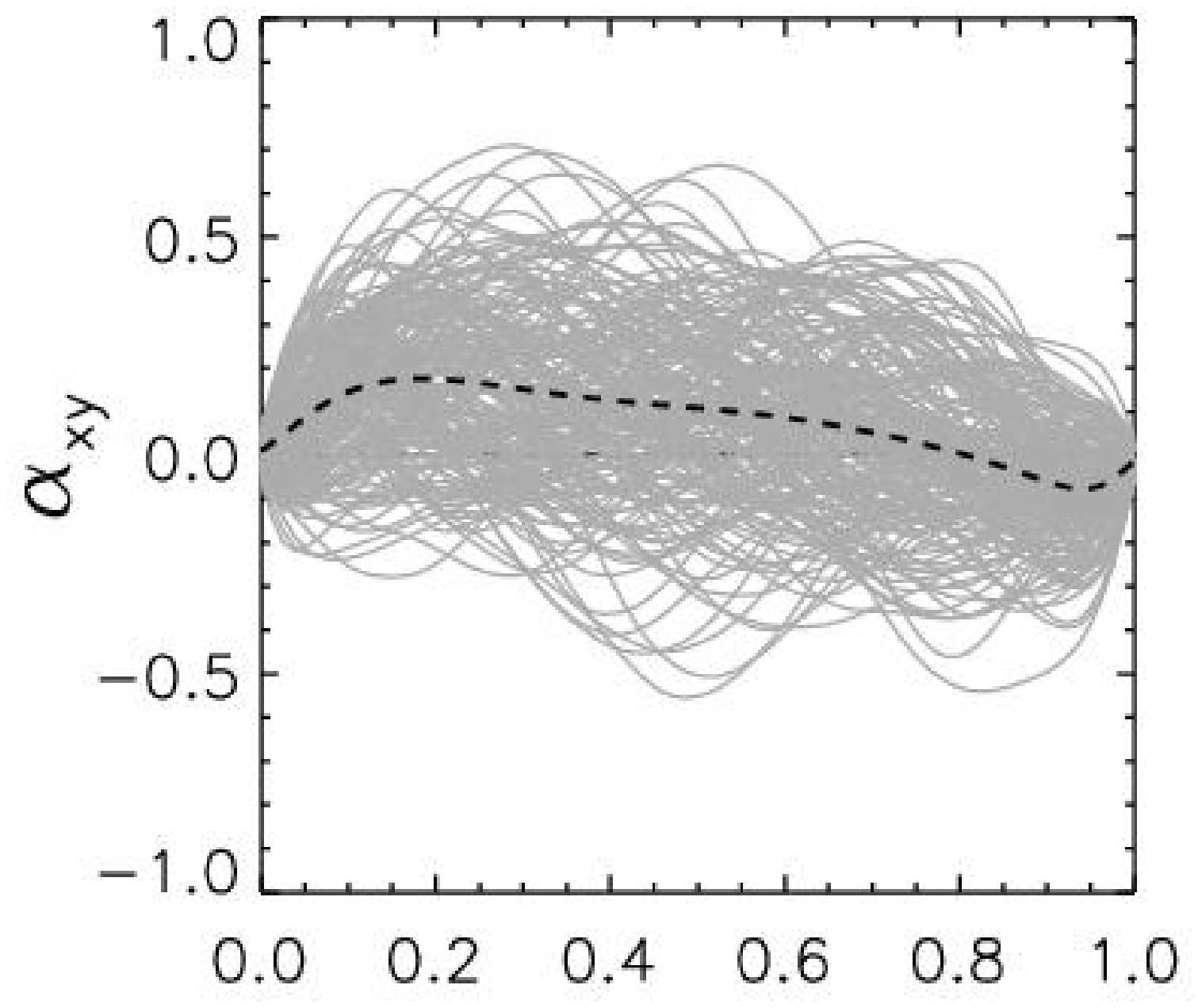}
\includegraphics[width=3.4cm]{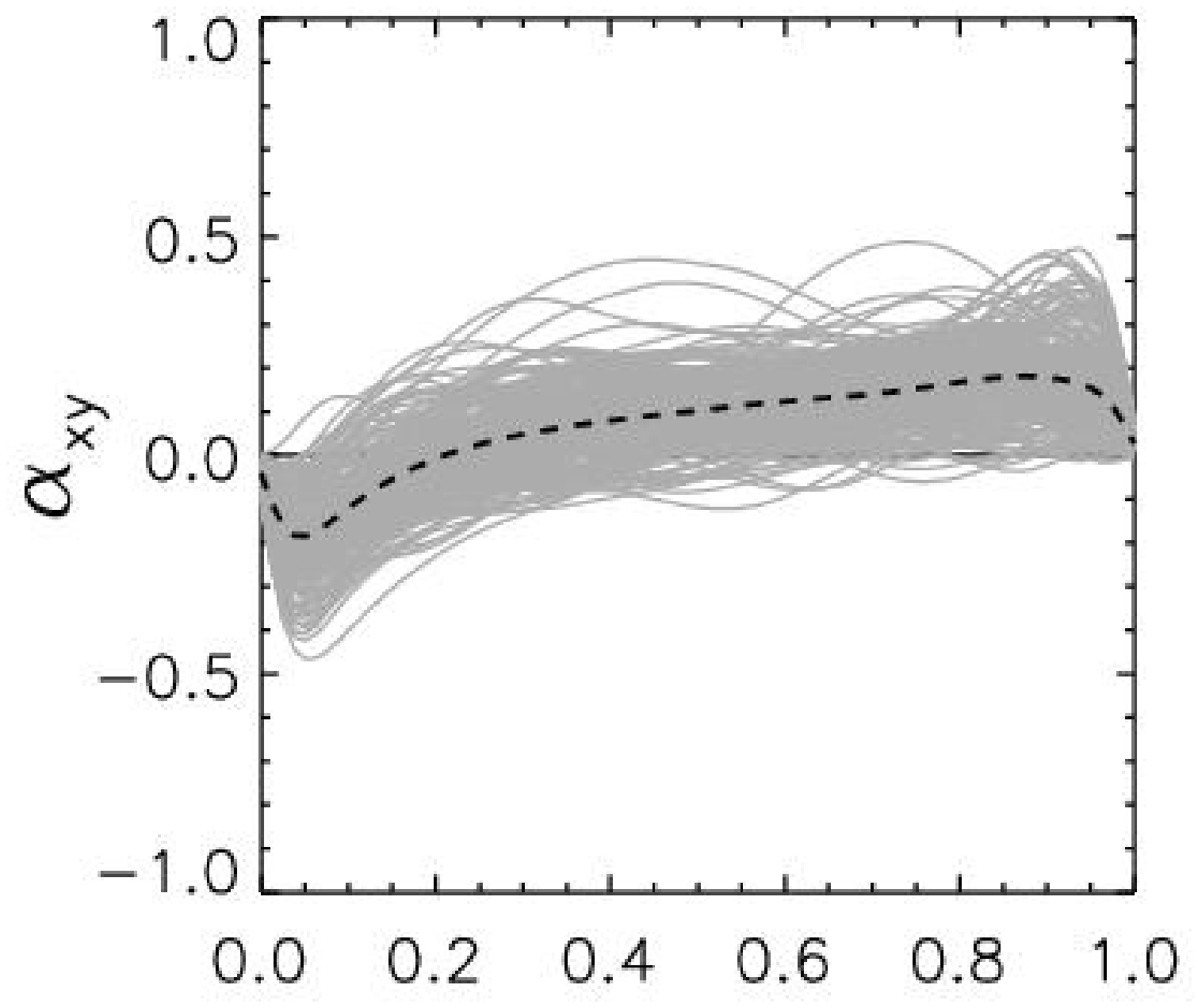}
\includegraphics[width=3.4cm]{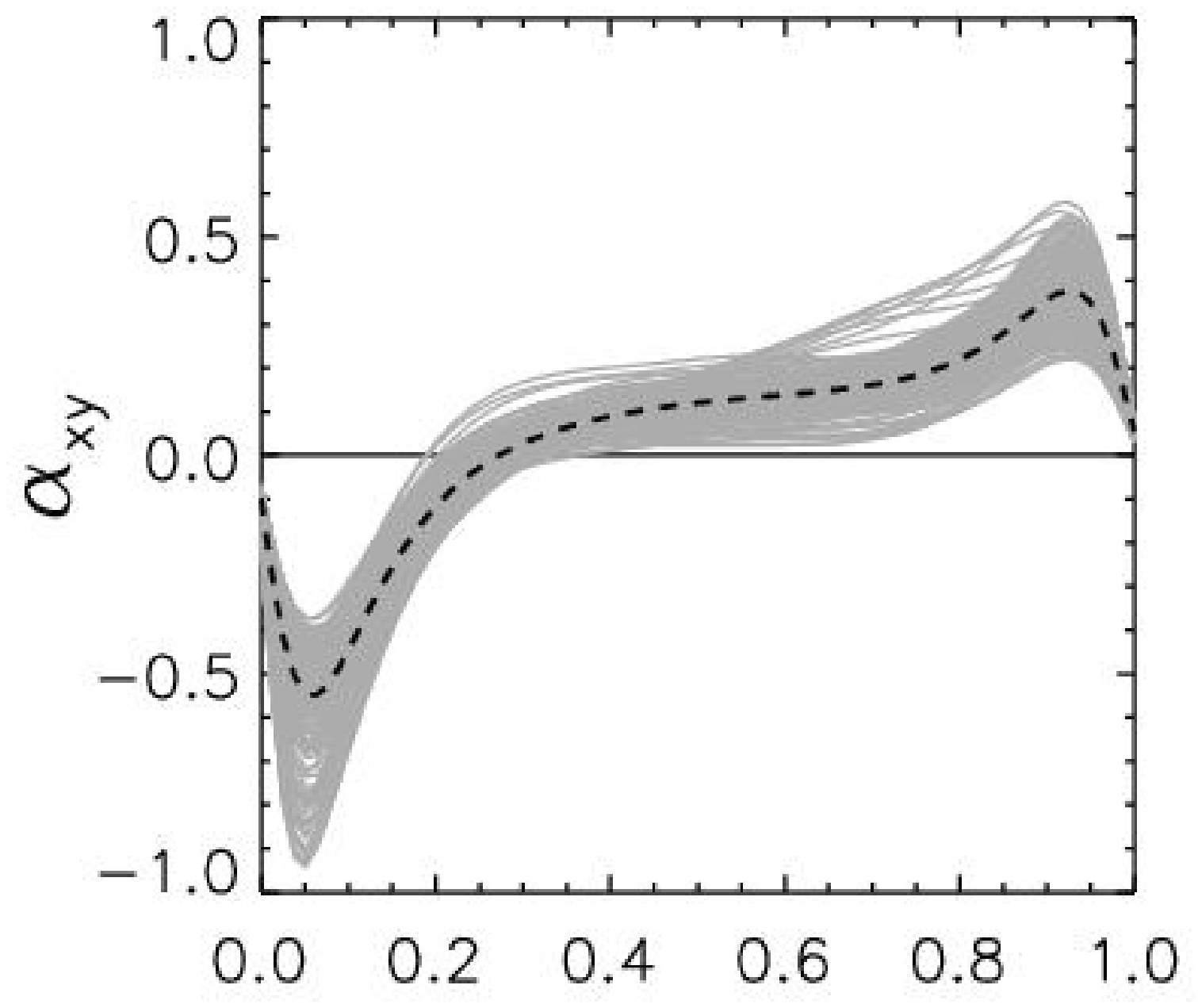}
\includegraphics[width=3.4cm]{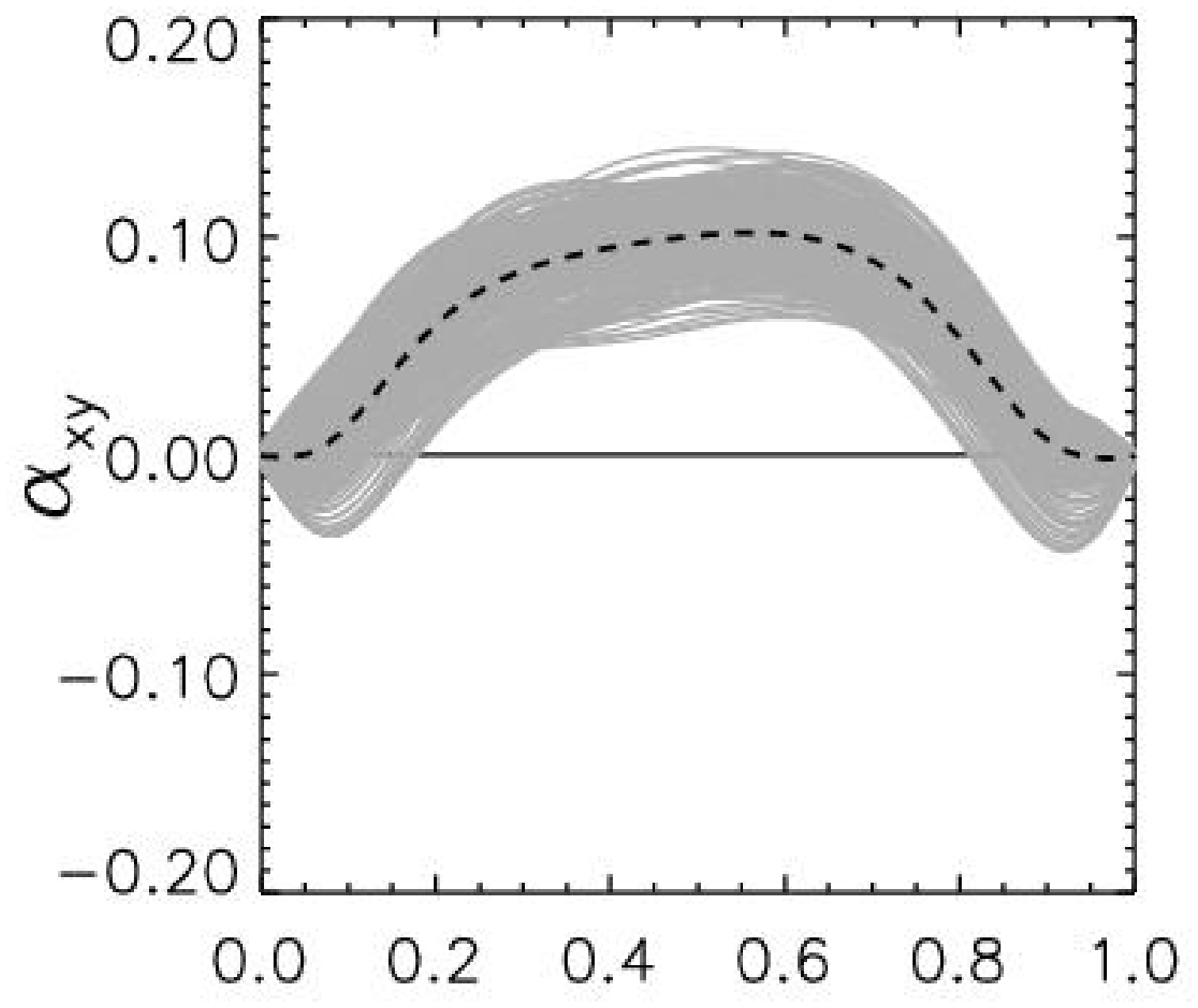}
\\
\includegraphics[width=3.4cm]{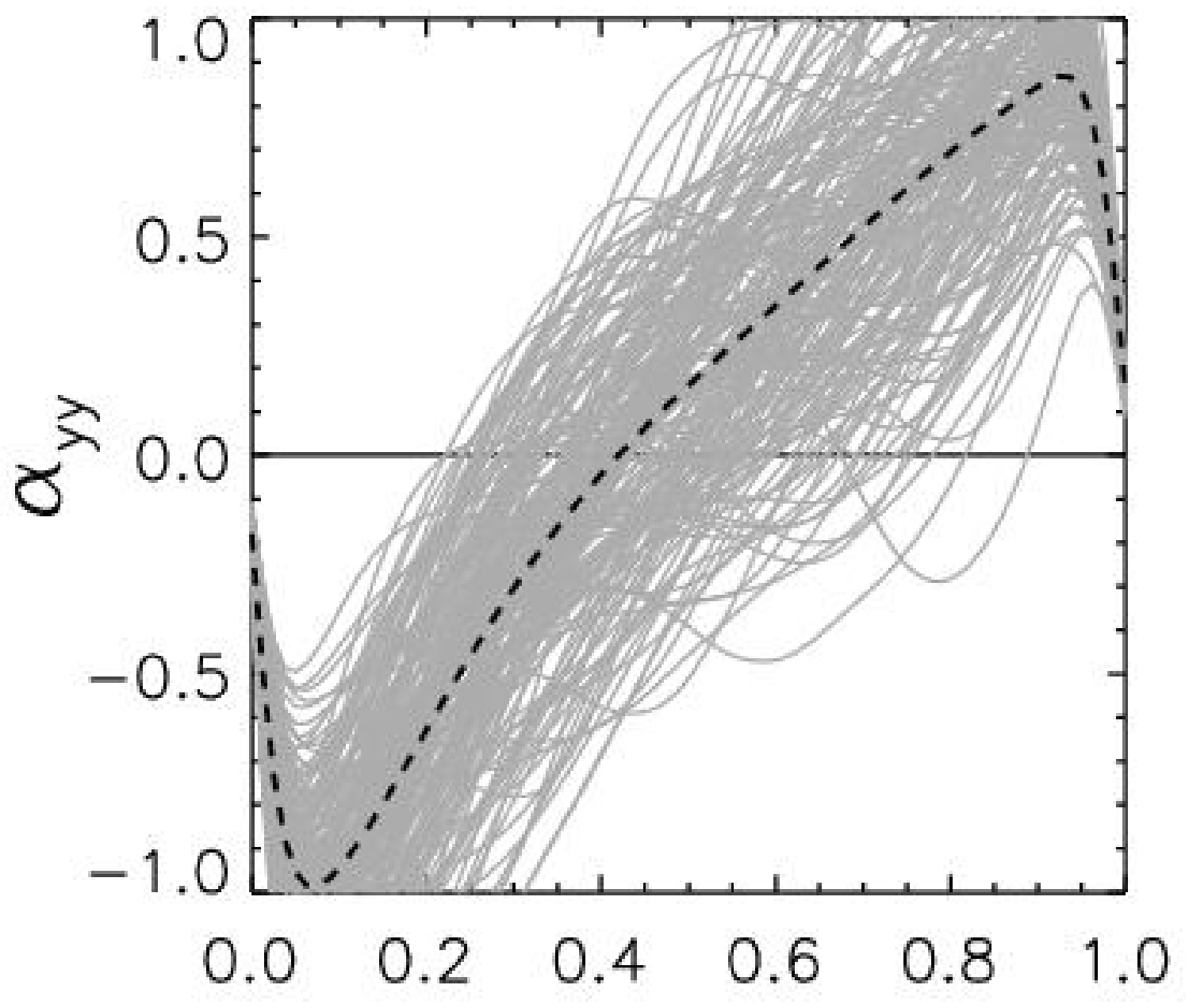}
\includegraphics[width=3.4cm]{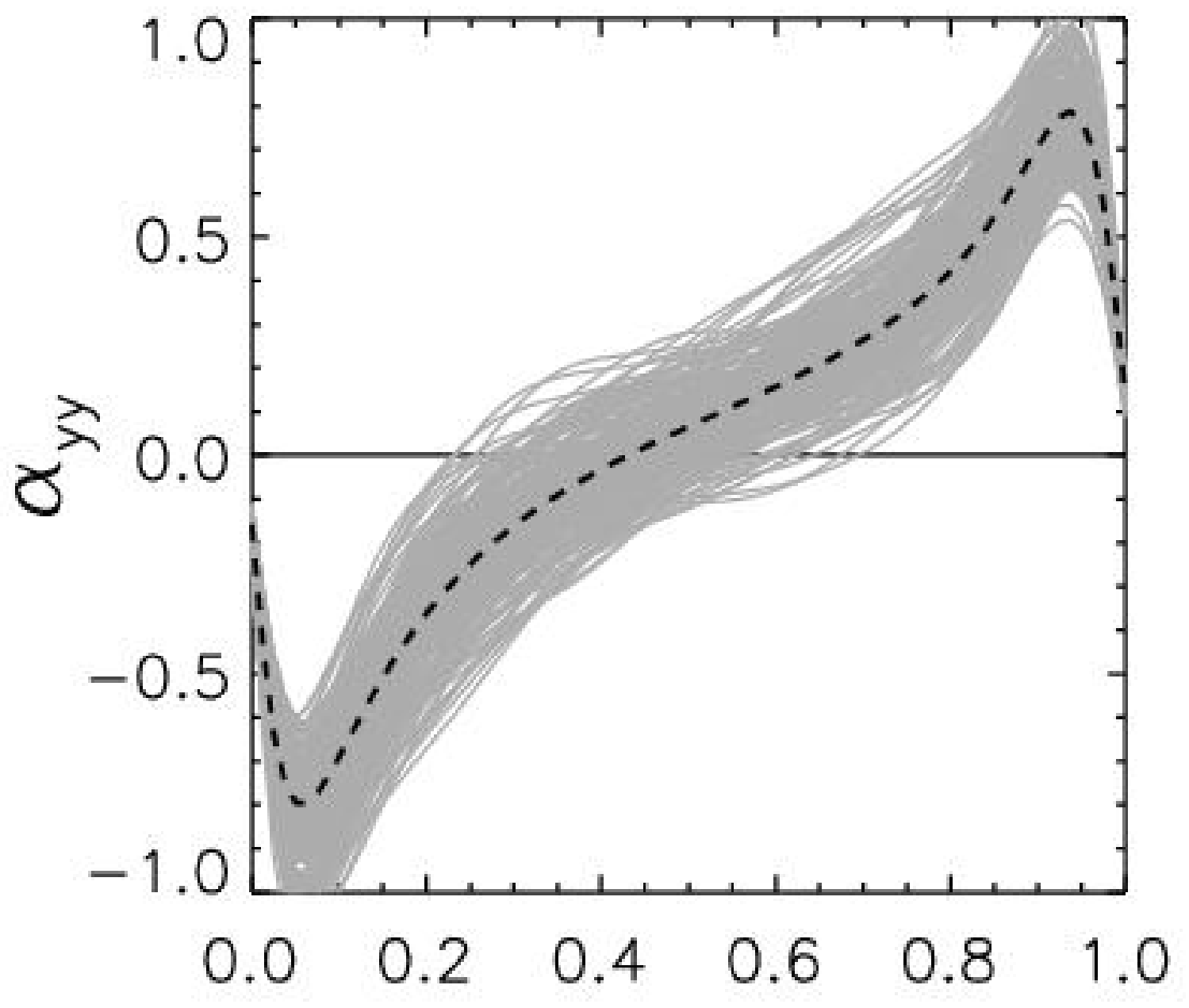}
\includegraphics[width=3.4cm]{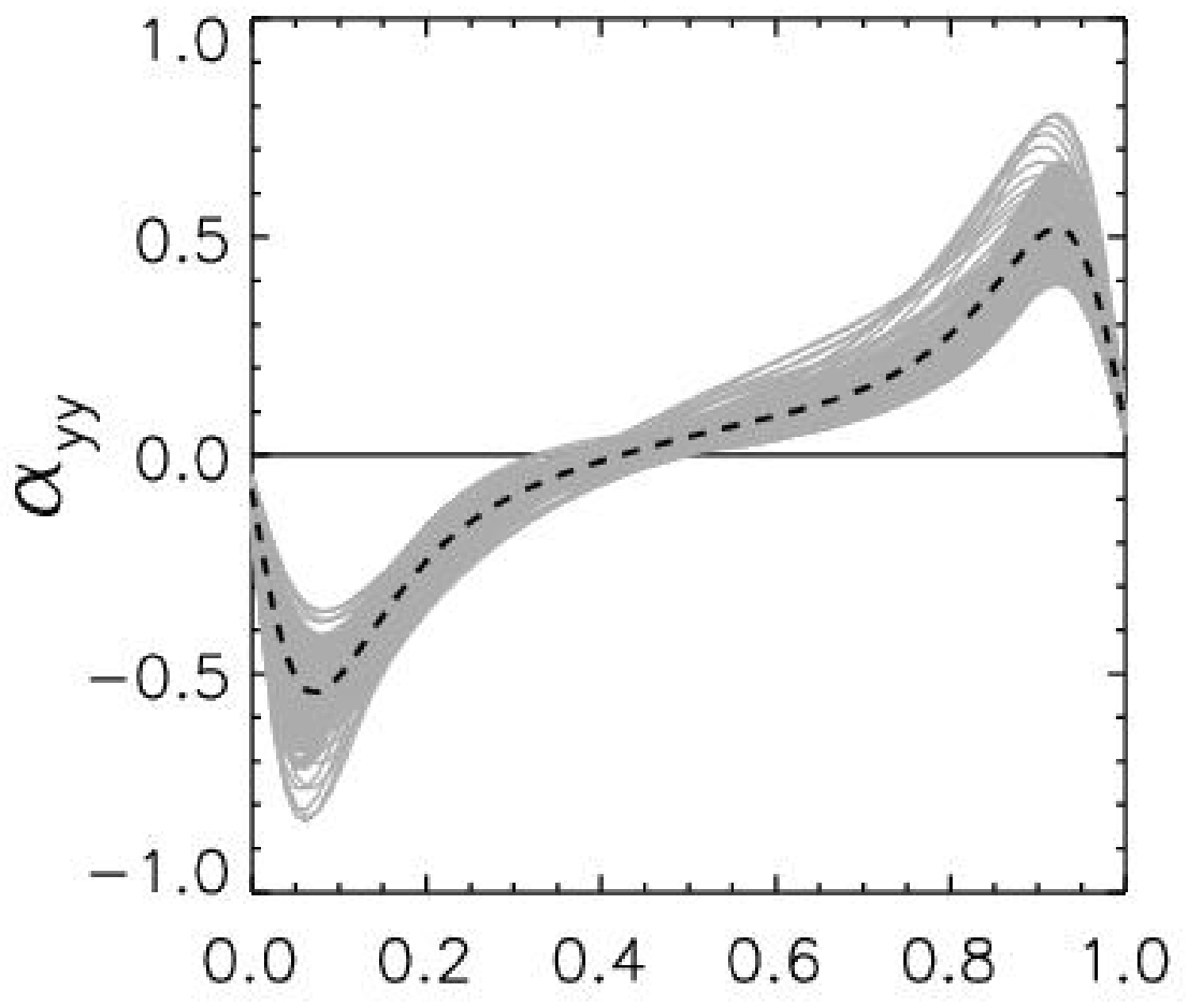}
\includegraphics[width=3.4cm]{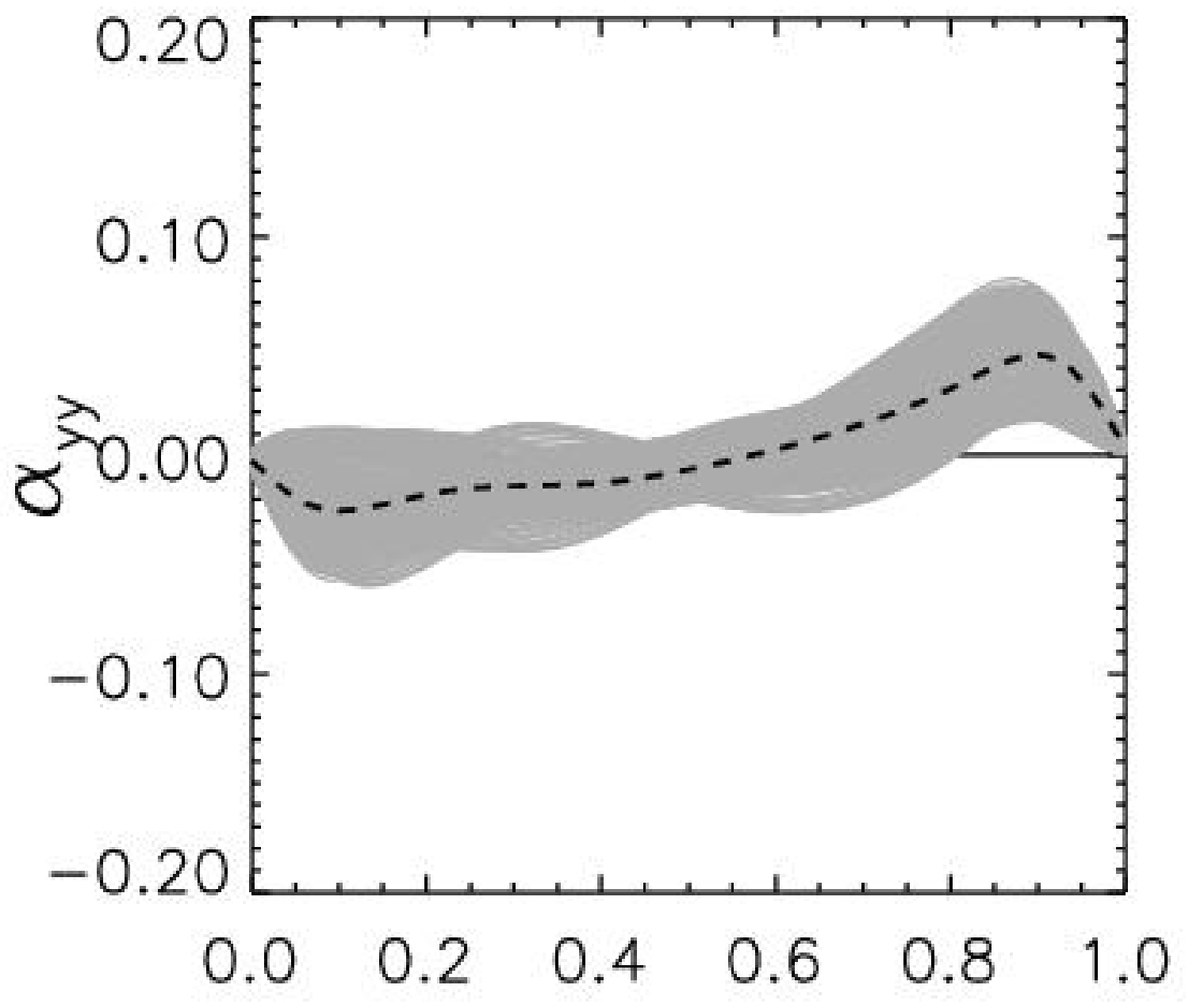}
\caption{$z$-dependence of $\alpha_{xy}$ (top)
  and $\alpha_{yy}$ (bottom) for $\Lambda = 0.1, 1, 10,
  100$ (from left to right) at $\theta = 45^\circ$. Note the reduction in
  scale by a factor of 5 for the $\Lambda=100$ case.}
\label{alpha_plot}
\end{figure}

In principle, with the present simulations it is possible to calculate the
three coefficients $\alpha_{xy},\alpha_{yy}\mbox{ and }\alpha_{zy}$.
Here, we want to concentrate on the first two coefficients
$\alpha_{xy}$ and $\alpha_{yy}$. 
$\alpha_{xy}$ describes turbulent (radial) advection of magnetic field (pumping), whereas
$\alpha_{yy}$ describes the production of magnetic field perpendicular to
$\left<B_y\right>$ via the $\alpha$-effect and is therefore of profound interest.  
The vertical profile of the calculated $\alpha$-coefficients
is plotted in Figure~\ref{alpha_plot} for $\Lambda=0.1,1,10,100$.
 The top row shows $\alpha_{xy}$ and the bottom row shows $\alpha_{yy}$. 
Again the grey lines represent snapshots of the coefficients at different times and the dashed line 
in each plot
represents the time-averaged $z$-profile. 

We note that $\alpha_{yy}(z)$ shows a quite asymmetric behavior with respect to $z$.
It is always negative in the lower part of
the layer and positive in the upper part of the layer.
The transition between negative and positive $\alpha$-effect occurs roughly in
 the middle of the box.
The peaks of $\alpha_{yy}$ are remarkably broad leading to two extended zones with
positive and negative $\alpha$-effect having nearly equal amplitude.
Comparable solutions proportional to $\sin(2\pi z)$ from quasi-linear
calculations have been found by \cite{1979PEPI...20..134S}.
For increasing magnetic field time fluctuations in the
$\alpha$-coefficients are obviously reduced which comes from the fact that the
flow tends to become laminar for large $\Lambda$.
The resulting vertical profiles of $\alpha_{yy}$ differ 
from computations with stronger stratification where the $z$-dependence is
much more asymmetric.
In the latter scenario the zero line is crossed in the lower half-box and 
the peak of
the $\alpha$-coefficients is higher in the upper half-box.
This leads to a non-zero $\alpha$-effect when averaged
over the entire box volume \citep{2004book_rued&holl}.
In contrast to this, the volume averaged $\alpha$-effect is rather small in
our model.

$\alpha_{xy}(z)$ shows a drastic change in its behavior between $\Lambda=0.1$ and $\Lambda=1$.
For $\Lambda=0.1$ $\alpha_{xy}$ has a broad peak located in the lower part
of the layer and changes its sign close to the upper vertical boundary.
For $\Lambda\geq 1$ $\alpha_{xy}$ shows a reversed $z$-dependence.
This feature is correlated with the change of behavior of
$\left<B_y\right>$ at the boundaries between $\Lambda=0.1$ and $\Lambda=1$.

\begin{figure}[h]
$\hspace*{2.8cm}\Lambda=1\hspace*{6cm}\Lambda=10$\\
\includegraphics[width=3.4cm]{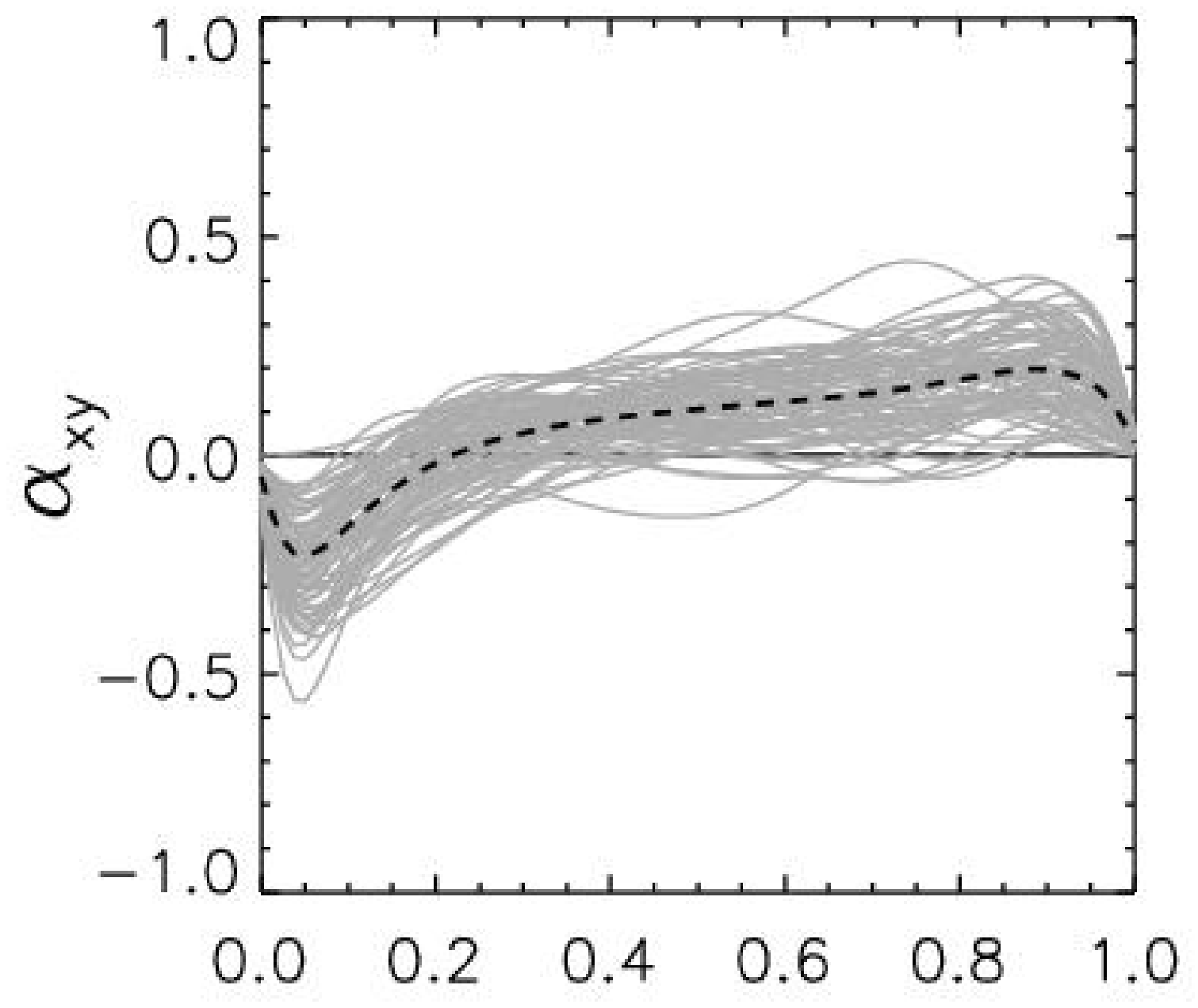}
\includegraphics[width=3.4cm]{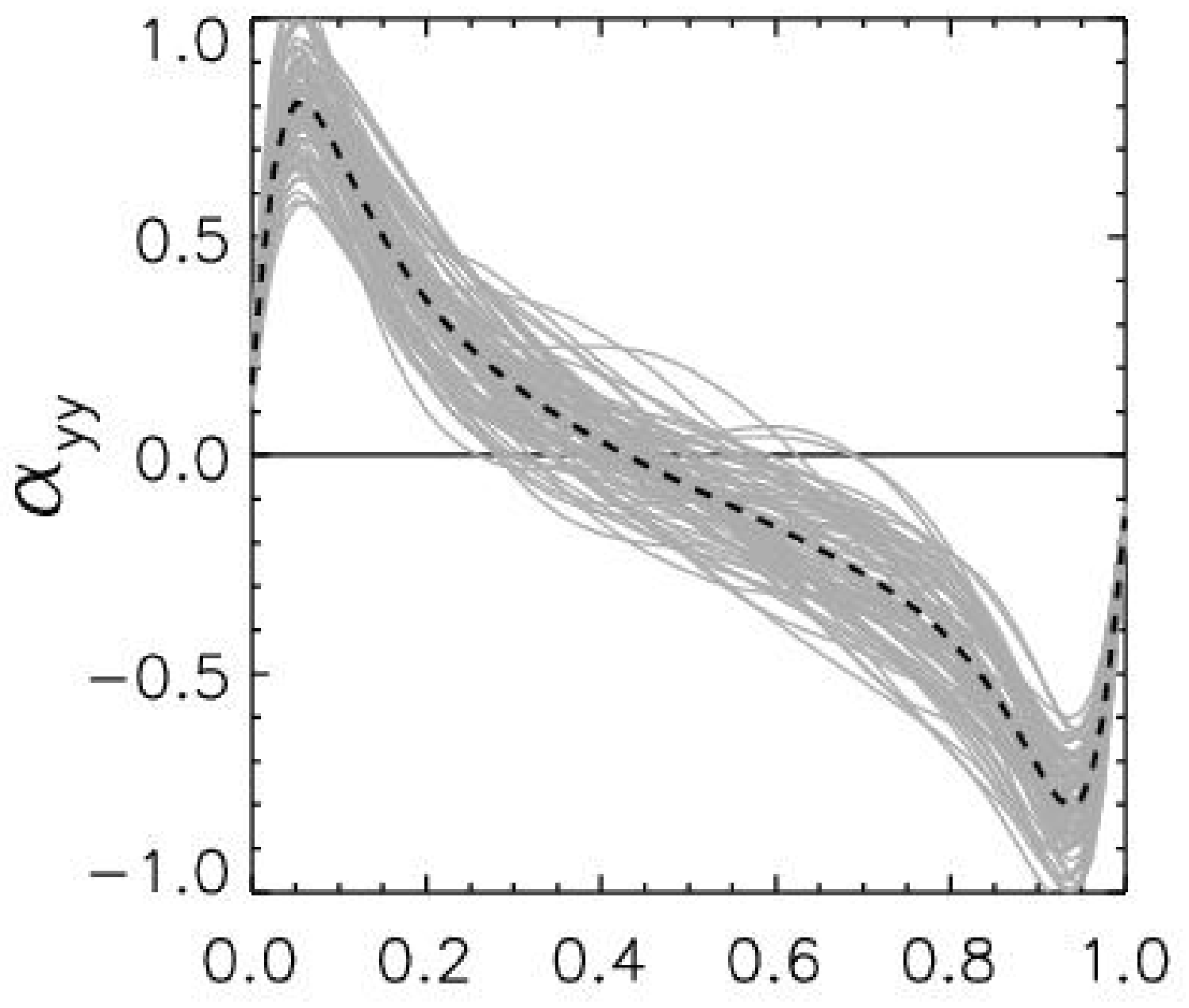}
\includegraphics[width=3.4cm]{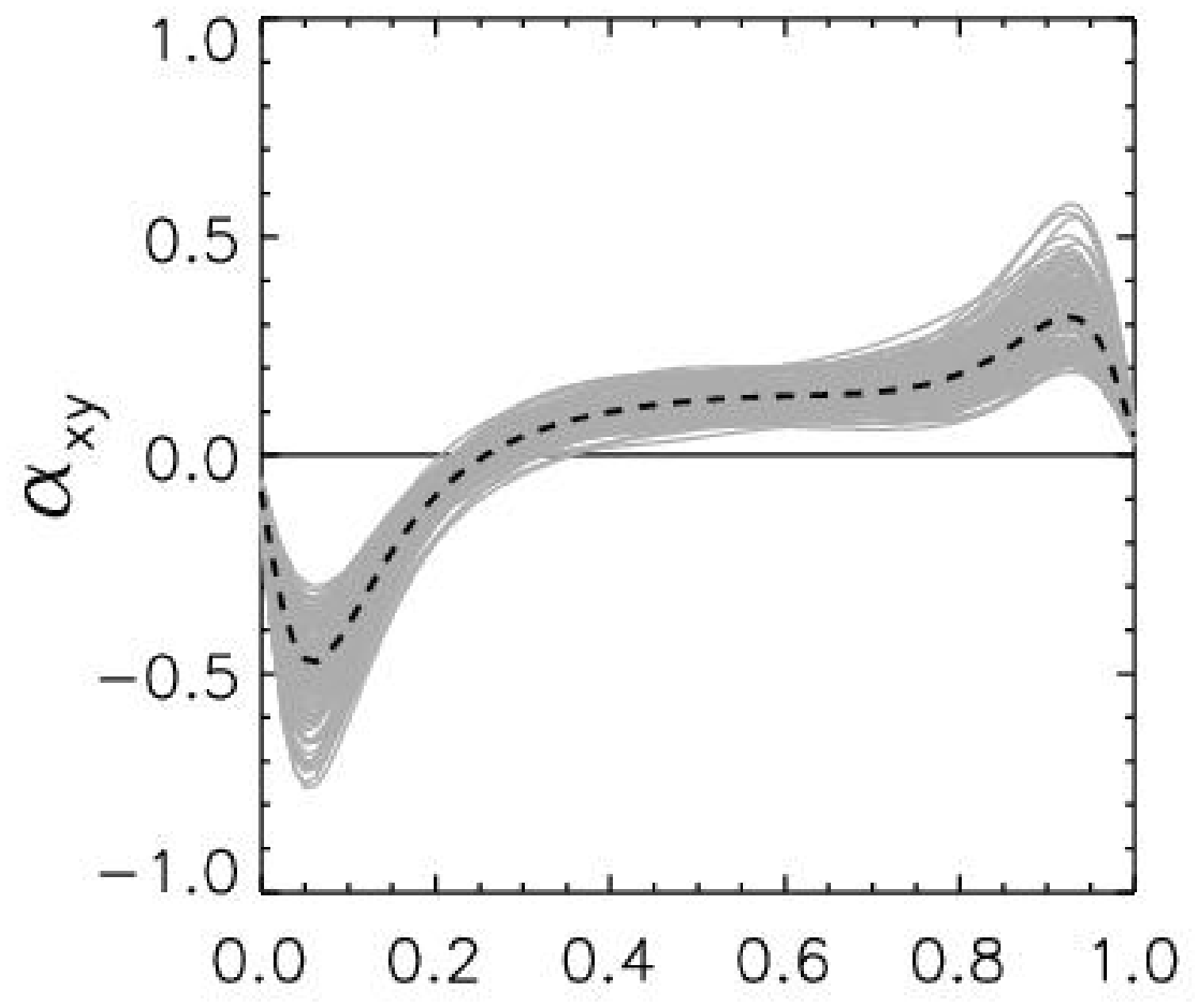}
\includegraphics[width=3.4cm]{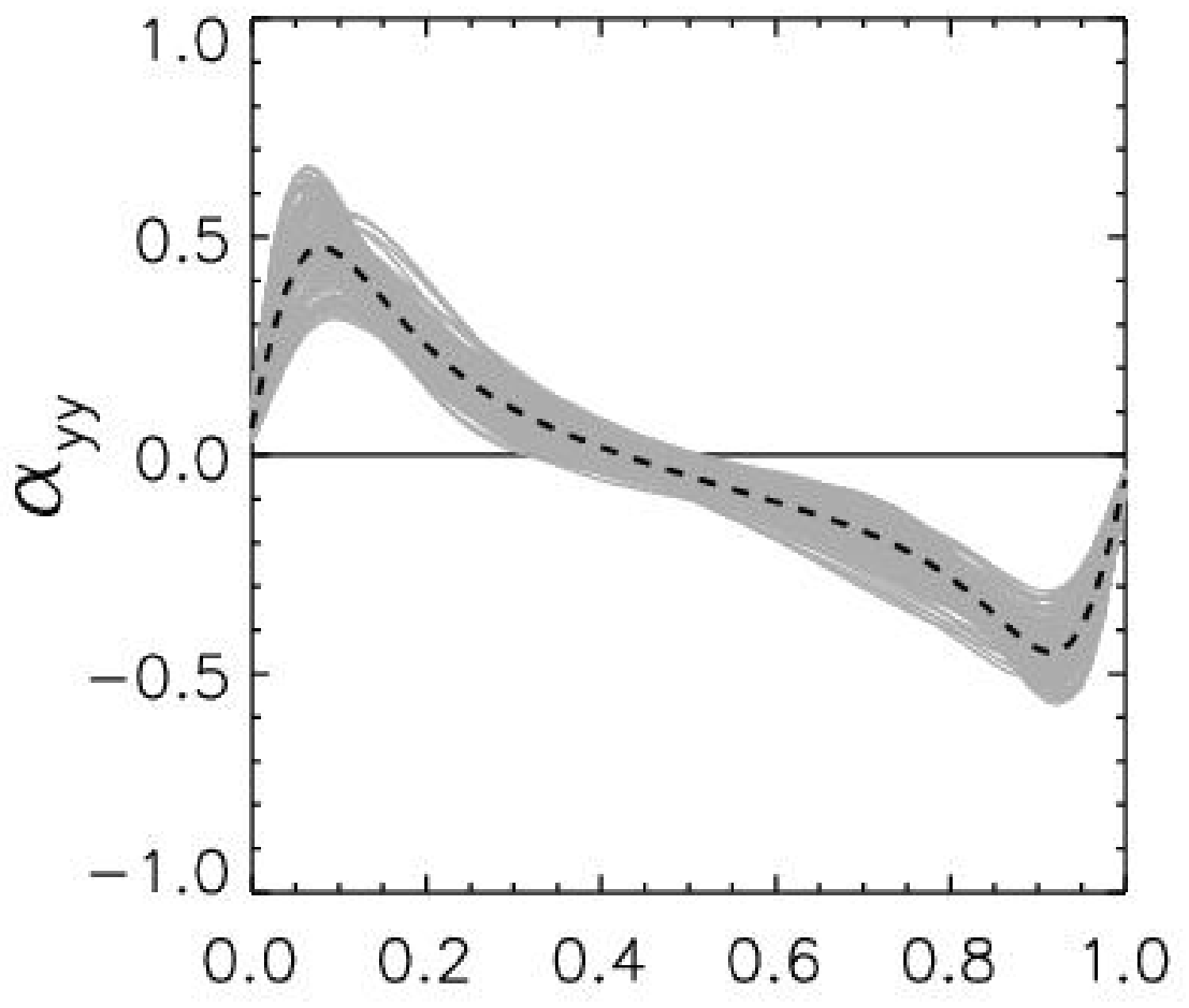}
\\
\includegraphics[width=3.4cm]{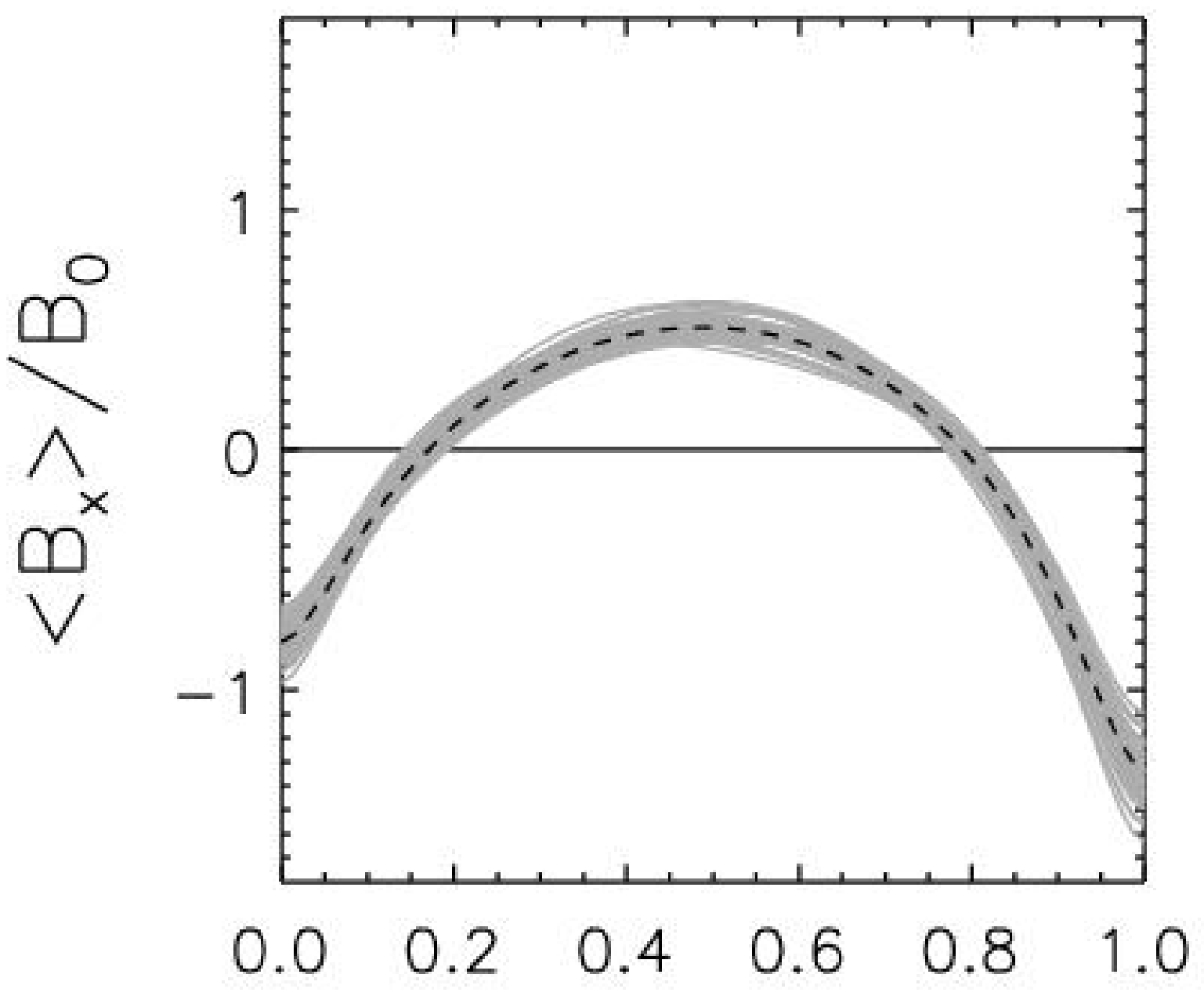}
\includegraphics[width=3.4cm]{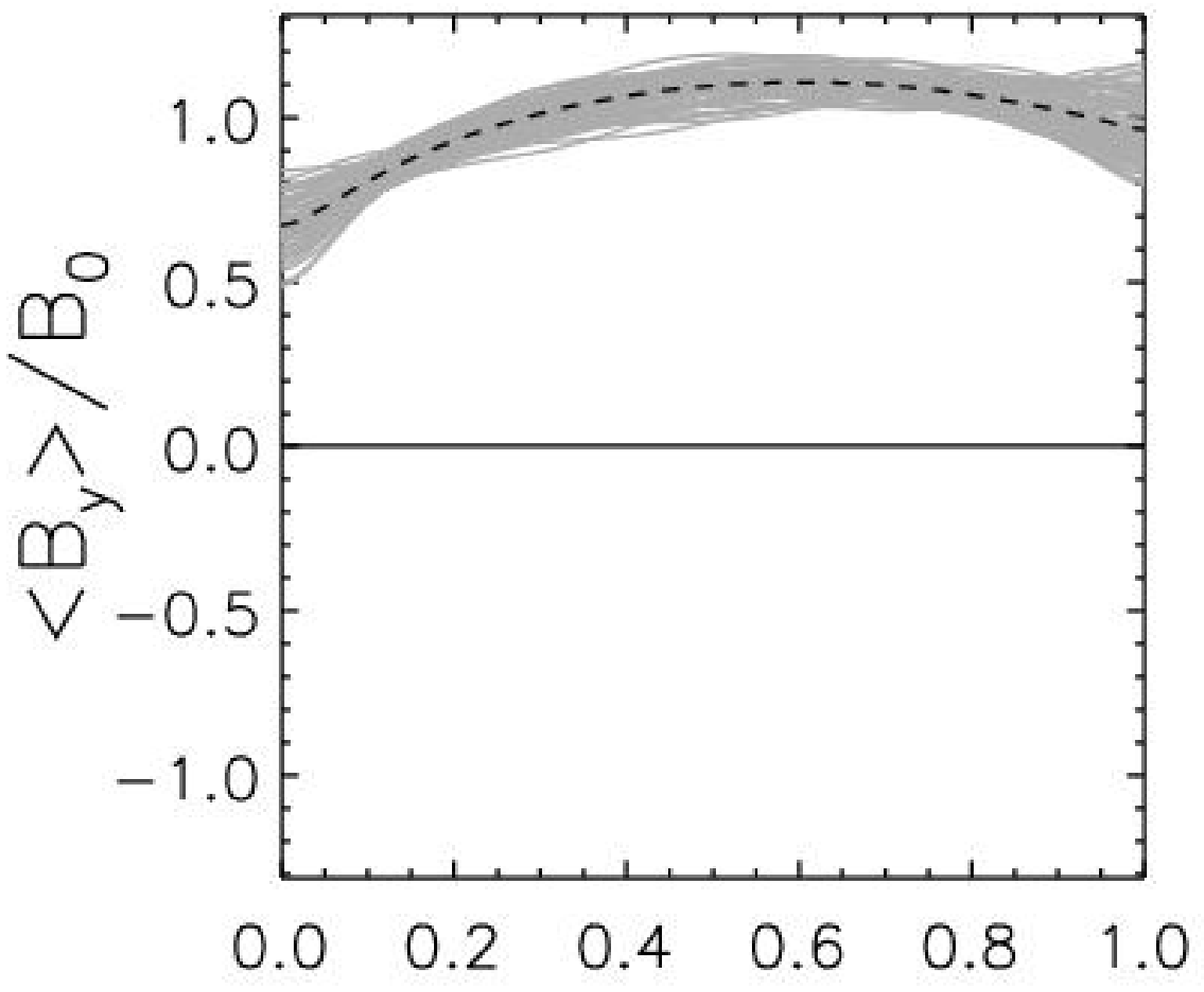}
\includegraphics[width=3.4cm]{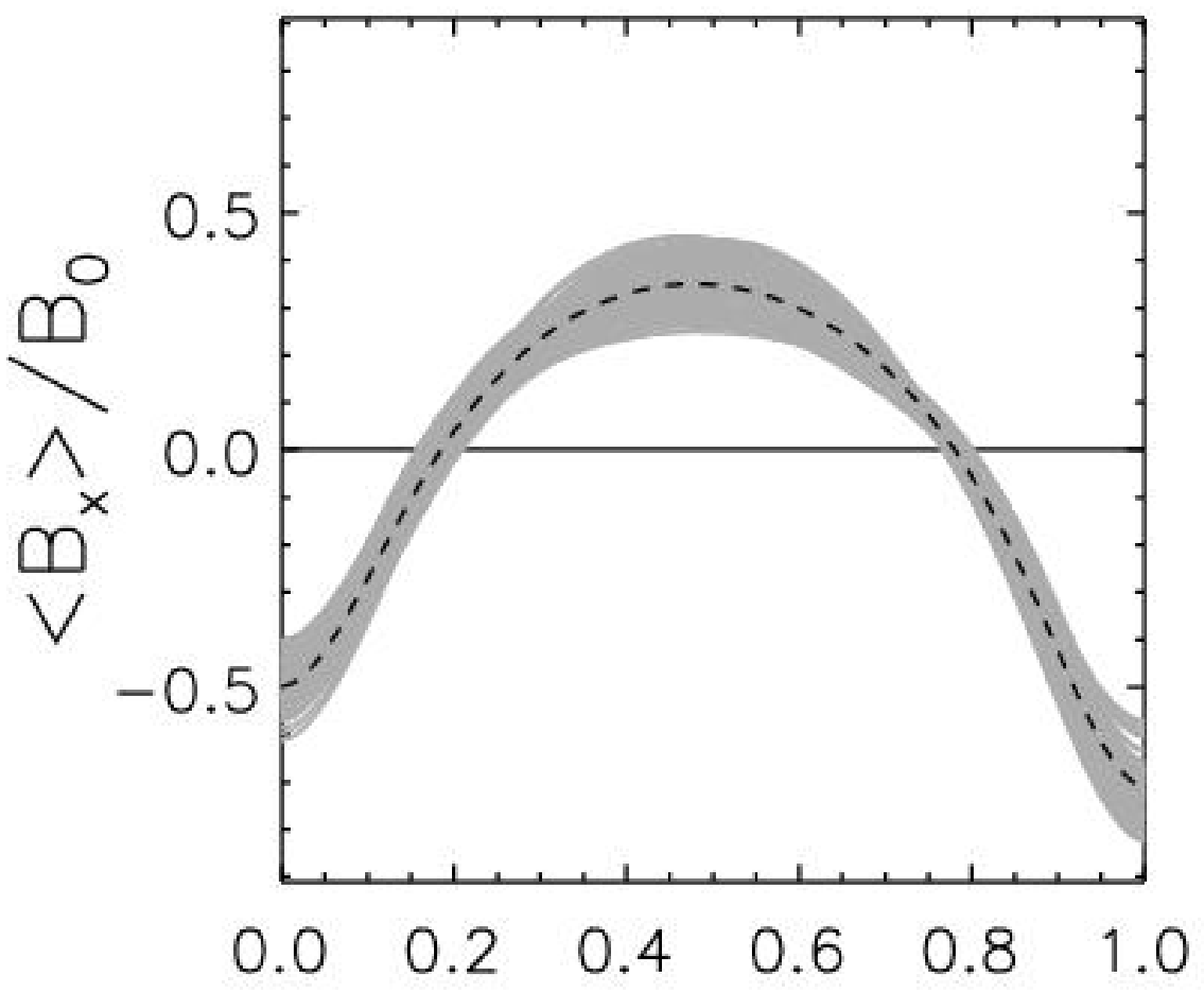}
\includegraphics[width=3.4cm]{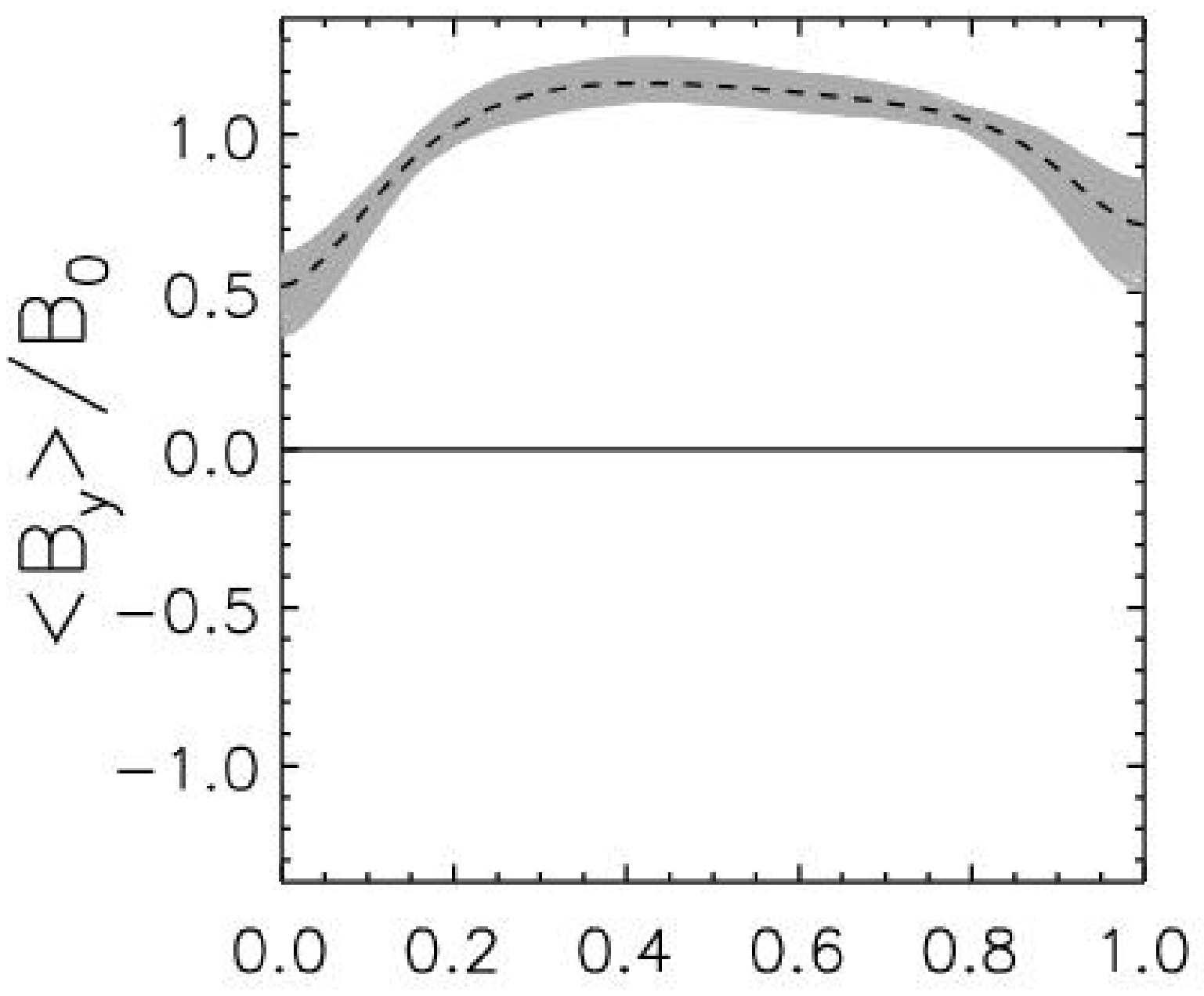}
\caption{$z$-dependence of $\alpha_{xy}$
  and $\alpha_{yy}$ (upper row) and the corresponding $\left<B_x\right>$
  and $\left<B_y\right>$ (lower row) for $\Lambda = 1$ and $\Lambda=10$ at $\theta = 135^\circ$}
\label{symetry}
\end{figure}

Computations with a box located at $\theta = 135^\circ$ reveal $\alpha_{xy}$ 
symmetric with respect to the equator but $\alpha_{yy}$ antisymmetric (see Figure~\ref{symetry}). 
The resulting $\left<B\right>(z)$ closely resemble the symmetry properties of
the $\alpha$-coefficients ($\left<B_x\right>$ antisymmetric,
$\left<B_y\right>$ symmetric). 
From that it follows that the coefficient $\alpha_{xy}$ plays the role of an (negative) advection velocity ($u^{\rm esc} = - \alpha_{xy}$).
The $z$-profile of $\alpha_{xy}$ shows no preferred symmetry with respect to the middle of the unstable
layer, and $\alpha_{xy}$ is -- independently of $\Lambda$ --
predominantly positive.
The escape velocity 
in the vertical direction is thus directed {\em downwards} for strong magnetic fields. 
Exactly this behavior has been obtained in a quasi-linear approximation by \cite{1992A&A...260..494K}
characterized as 'turbulent buoyancy'. 
The same result has been found in numerical simulations by \cite{2001A&A...365..562D}, \cite{2002A&A...394..735O} and \cite{2003A&A...401..433Z}. It can now be considered as a well-established phenomenon that the magnetic-induced turbulent pumping 
transports mean magnetic field downwards rather than upwards to the surface.
\subsection{Kinetic helicity}
\begin{figure}[h]
\includegraphics[width=3.4cm]{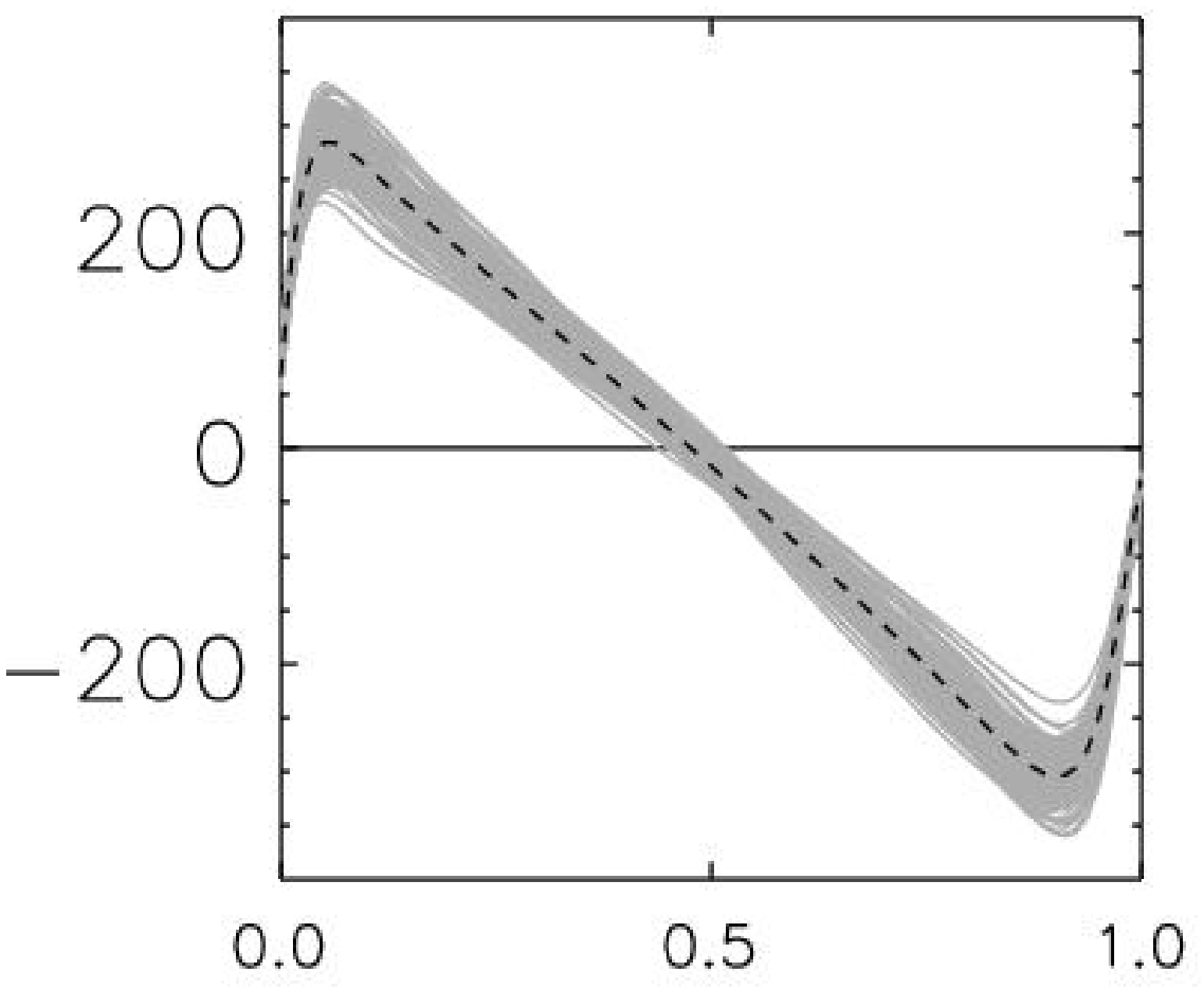}
\includegraphics[width=3.4cm]{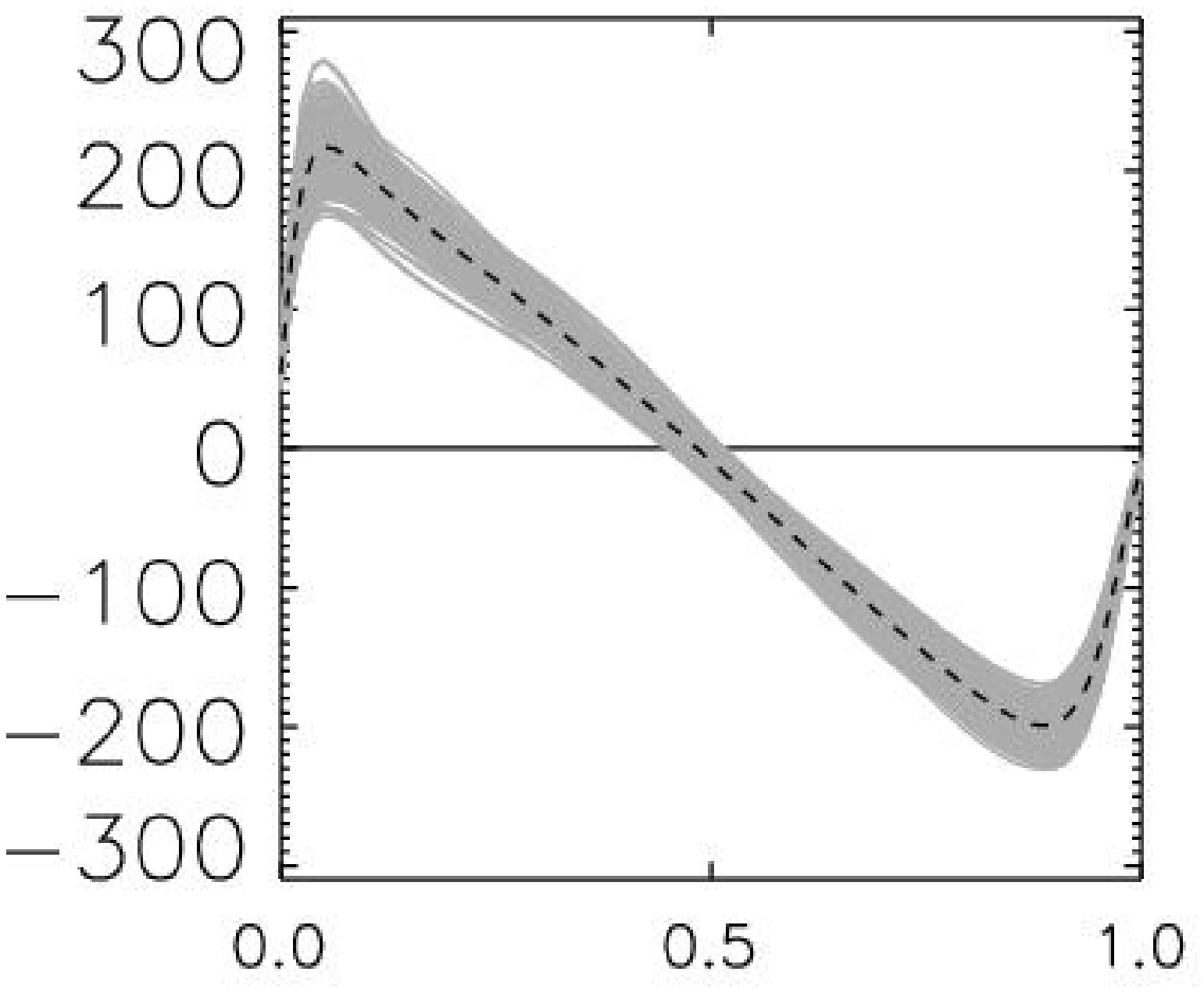}
\includegraphics[width=3.4cm]{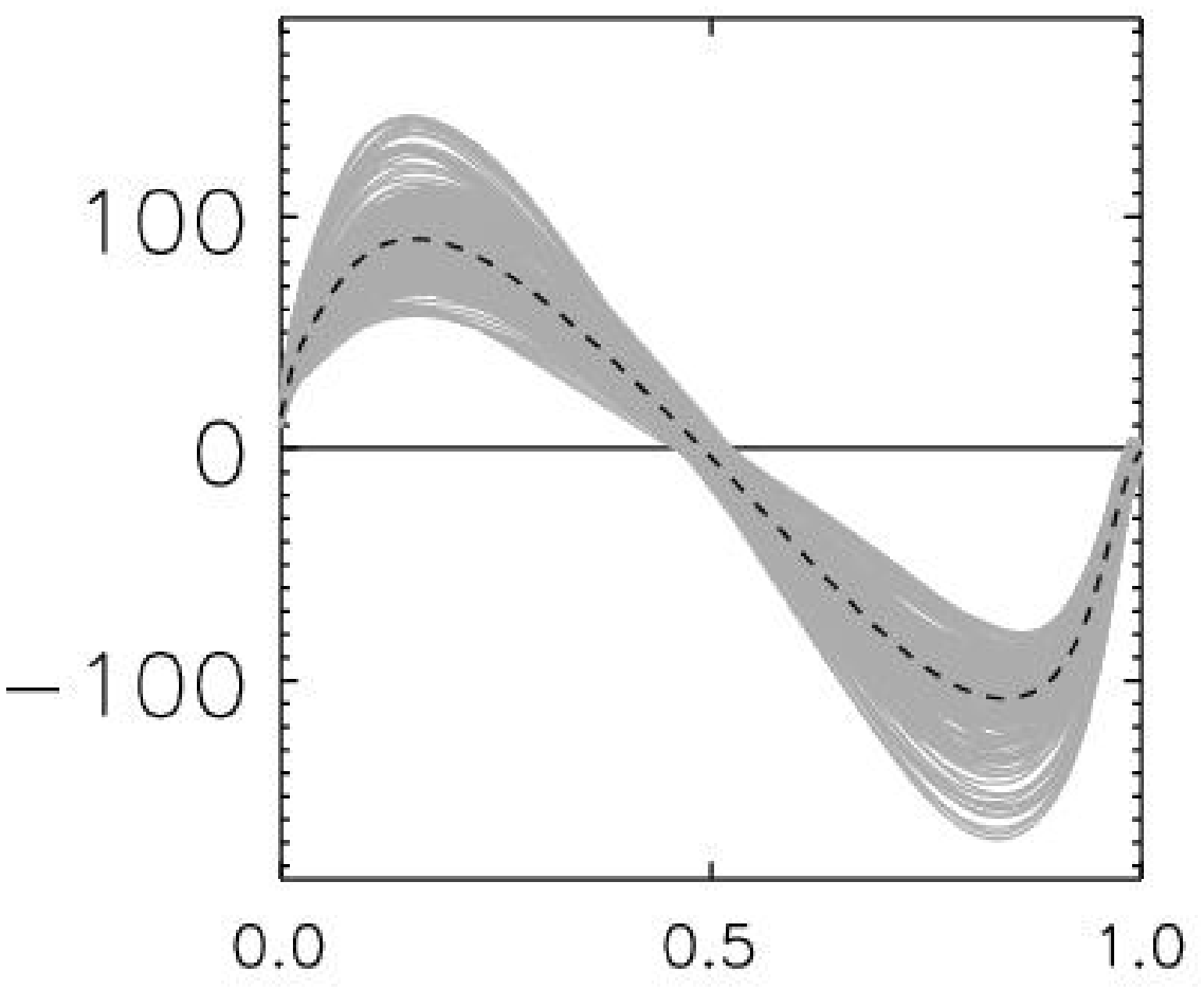}
\includegraphics[width=3.4cm]{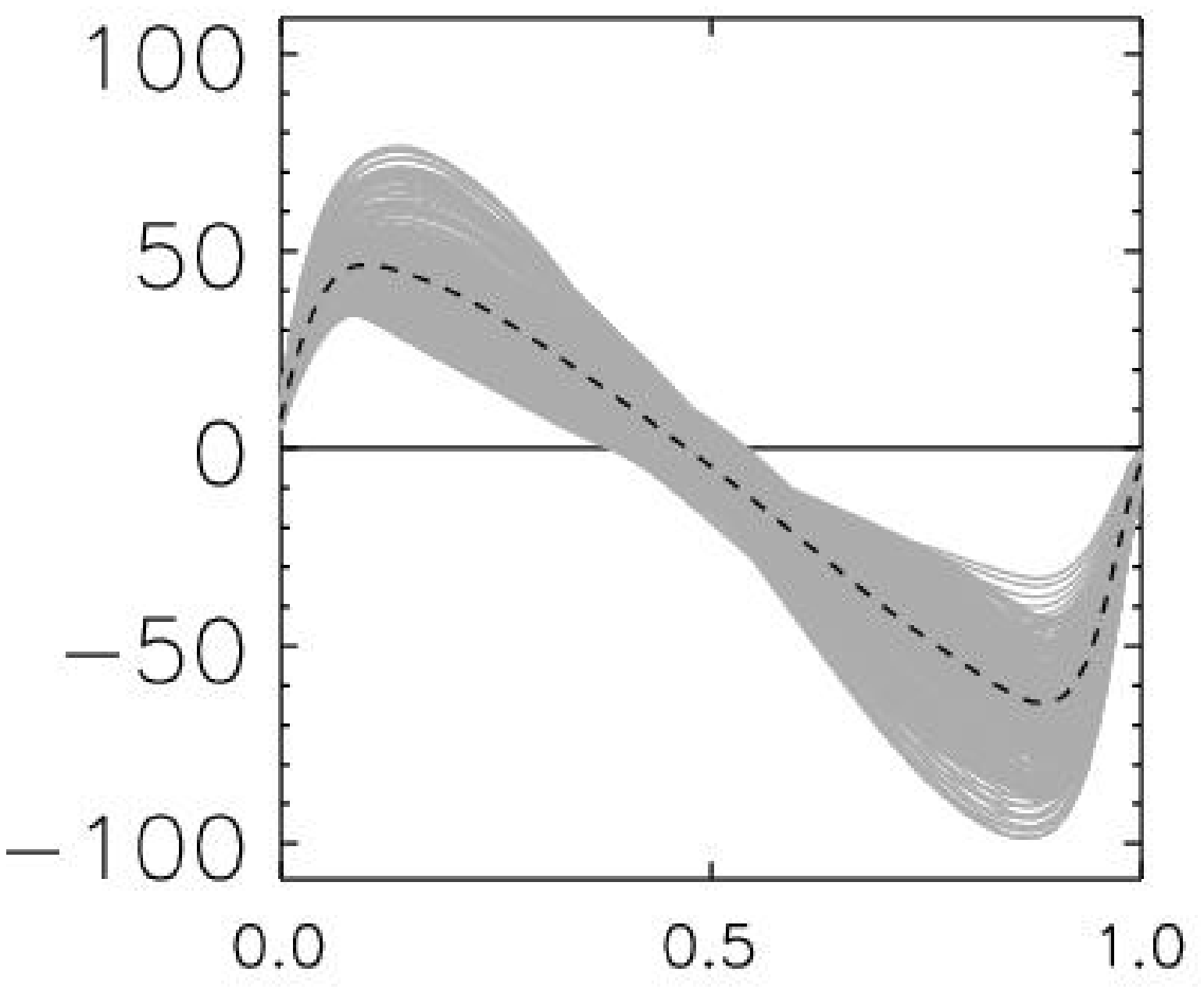}
\caption{$z$-dependence of the kinetic helicity for $\Lambda=0.1,1,10,100$}
\label{helicity_plot}
\end{figure}

Figure~\ref{helicity_plot} shows the $z$-profile of the kinetic helicity
${H}_{\mathrm{kin}}=\left<\vec{u}'\cdot\nabla\times{\vec{u}'}\right>$.
The kinetic helicity is negative in the upper part of the box and positive in
the lower part. 
Comparing figure~\ref{helicity_plot} with the corresponding $\alpha_{yy}$-profiles from figure~\ref{alpha_plot}
the well-known relation between the signs of the $\alpha$-coefficient
and the kinetic helicity is confirmed, i.e.
$\alpha\sim-1/3\tau_{\mathrm{cor}}\left<\vec{u}'\cdot\nabla\times{\vec{u}'}\right>$ with a correlation
time $\tau_{\mathrm{cor}}$(see e.g. \citeauthor*{1980mfmd.book.....K}, \citeyear{1980mfmd.book.....K}).
A similar relation 
was found by \cite{2001A&A...376..713O}. 
Increasing the strength of the imposed
magnetic field the change in amplitude of the helicity is roughly in
accordance with the change in amplitude
of $\alpha_{yy}$.
The correlation time $\tau_{\mathrm{cor}}$ roughly is of the order of $10 \%$
of the turnover time $\tau_{\mathrm{adv}}$
%
%
%
%
%
%
%
\subsection{$\alpha$-quenching}
\begin{figure}[h]
\includegraphics[width=9cm]{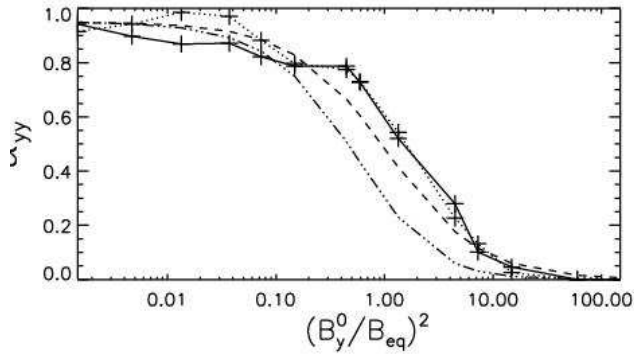}
\caption{Quenching behavior of
  $\alpha_{yy}$. A value of $(B_y^0/B_{\mathrm{eq}})^2=1$
  (equipartition value) corresponds to
  $\Lambda \approx 6.2$.  The solid (dotted) line corresponds to the maximum
  (minimum) of $\alpha_{yy}(z)$, the dashed line corresponds to the analytic
  quenching function (\ref{quench}) and the dashed-dotted line represents the
  analytic expression derived by \cite{1993A&A...269..581R}.
}
\label{quench_plot}
\end{figure}

In order to estimate the quenching behavior of the $\alpha$-effect we investigate the variation of the local maximum (minimum)
value of the time averaged $z$-profile $\alpha_{yy}(z)$ in dependence of the quantity
$(B_y^0/B_{\mathrm{eq}})^2$  where $B_y^0$ denotes the initially imposed magnetic
field strength and ${B}_{\mathrm{eq}}$ is the so
called equipartition field which is defined by
${B}_{\mathrm{eq}}^2=\mu_0\rho{u}_{\mathrm{rms}}^2({B}\!\rightarrow\! 0)$, i.e. $u_{\mathrm{rms}}$ is
related to the non-magnetic case.
In a rough approximation we have $\Lambda \approx 6.24\cdot (B_y^0/B_{\mathrm{eq}})^2$.
The quenching curves are plotted in figure~\ref{quench_plot} 
where the solid (dotted) line corresponds to the maximum (minimum) of $\alpha_{yy}(z)$. 
The dashed line corresponds to a simple analytic quenching function of the form
\begin{equation}
\alpha({B})=\alpha({B}\!\rightarrow\! 0)\cdot \frac{1}{1+\left(B_y^0/B_{\mathrm{eq}}\right)^2}.
\label{quench}
\end{equation}
A detailed analysis of the $\alpha$-quenching phenomenon for slow rotation,
however, can
be found in \cite{1993A&A...269..581R}.
The corresponding expressions for $\alpha_{yy}$ is given by the dashed-dotted
line in figure \ref{quench_plot}. 
It overestimates the suppression of $\alpha_{yy}$
so that $\alpha$-quenching is probably better described by equation
(\ref{quench}) for our case of a fast rotator. 
Note that the curves associated with the maximum respective minimum of
$\alpha_{yy}$ are almost identical for $(B_y^0/B_{\mathrm{eq}})^2
\gsim  0.2$ and
the deviation is at most about $10\%$ for $(B_y^0/B_{\mathrm{eq}})^2 \lsim 0.2$.  

For the dynamo number,  $C_{\mathrm{\alpha}}=\alpha d/\eta$, we obtain a value of the order of
$10$ for imposed fields below $(B_y^0/B_{\mathrm{eq}})^2\approx 1$.
This value might be large enough to allow global dynamo action in the sense of
$\alpha^2$-dynamos but this, of course, is  a very crude estimation. 
%
%
%
%
%
%
\section{Conclusions}
The main result of our computations is that the $y$-component of the EMF
describing the $\alpha$-effect in the azimuthal direction
is positive in the upper
part of the unstable layer and negative in the lower part on the northern hemisphere.
We obtained the $z$-profiles of the $\alpha$-coefficient for a wide range of imposed magnetic
field strength with the amplitude of the $\alpha$-effect probably sufficient to ensure dynamo
action in  mean-field calculations. 

The $\alpha$-effect is quenched
 under the influence of  strong external  magnetic fields but the suppression
 is not catastrophic, and significant quenching only sets in for magnetic
 fields above the equipartition value ${B_{\mathrm{eq}}}$.

Some weak points in the above model calculations should be mentioned.
First, we neglected the radial dependence of the gravitational force and the
effects of spherical geometry, thus, the influence of curvature effects remains unknown. 
Compositional
convection is ignored and the adopted parameter values are still far from the realistic
values for the Earth. 
Somewhat unphysical boundary conditions have been used by assuming the conditions of a
perfect conductor for both vertical boundaries so it cannot be ruled
out that the obtained profiles are profoundly affected by the vertical boundary
conditions.
Test calculations with rigid boundary conditions for the
velocity and/or slightly different boundary conditions for the magnetic field
as  used e.g. by \cite{2002A&A...394..735O} or \cite{2002A&A...386..331Z} show,
however, no major
differences in the EMF profiles. 
It remains, therefore,  a much-promising 
 ansatz for future mean-field calculations of the geodynamo taking into
 account the special radial profile of the $\alpha$-effect as it has been obtained in the present paper.
\begin{ack}
{This work was supported by the DFG SPP 1097 ``Erdmagnetische Variationen:
  Raum-Zeitliche Struktur, Prozesse und Wirkungen auf das System Erde''.
The computations have been performed using the Hitachi SR8000 and the PC
Cluster at the Astrophysikalisches Institut Potsdam (AIP).}
\end{ack}
\bibliographystyle{elsart-harv}
\bibliography{references}
\end{document}